\documentclass[a4paper,11pt]{article}
\pdfoutput=1 

\usepackage{jheppub} 

\usepackage[T1]{fontenc} 

\usepackage{braket}
\usepackage{comment}

\title{\boldmath Extracting Hawking Radiation Near the Horizon of AdS Black Holes}

\author[a,b,c]{Krishan Saraswat}
\author[a,b,c]{and Niayesh Afshordi}

\affiliation[a]{Department of Physics and Astronomy, University of Waterloo, 200 University Ave W, Waterloo, Canada}
\affiliation[b]{Waterloo Centre for Astrophysics, University of Waterloo, Waterloo, ON, N2L 3G1, Canada}
\affiliation[c]{Perimeter Institute For Theoretical Physics, 31 Caroline St N, Waterloo, Canada}

\emailAdd{ksaraswat@pitp.ca}
\emailAdd{nafshordi@pitp.ca}

\abstract{We study how the evaporation rate of spherically symmetric black holes is affected through the extraction of radiation close to the horizon. We adopt a model of extraction that involves a perfectly absorptive screen placed close to the horizon and show that the evaporation rate can be changed depending on how close to the horizon the screen is placed. We apply our results to show that the scrambling time defined by the Hayden-Preskill decoding criterion,  which is derived in Pennington's work (arXiv:1905.08255) through entanglement wedge reconstruction is modified. The modifications appear as logarithmic corrections to Pennington's time scale which depend on where the absorptive screen is placed. By fixing the proper distance between the horizon and screen we show that for small AdS black holes the leading order term in the scrambling time is consistent with Pennington's scrambling time. However, for large AdS black holes the leading order Log contains the Bekenstein-Hawking entropy of a cell of characteristic length equal to the AdS radius rather than the entropy of the full horizon. Furthermore, using the correspondence between the radial null energy condition (NEC) and the holographic c-theorem, we argue that the screen cannot be arbitrarily close to the horizon. This leads to a holographic argument that black hole mining using a screen cannot significantly alter the lifetime of a black hole.}

\begin{document} 
\maketitle

\section{Introduction}
\label{introsec}
The AdS/CFT correspondence is a conjecture that relates gravitational systems in asymptotically AdS spacetimes to conformal field theories in one fewer spatial dimension \cite{Maldacena:1997re,Ramallo:2013bua,VanRaamsdonk:2016exw}. This provides an ideal setting to resolve the black hole information paradox \cite{PhysRevD.14.2460,Mathur:2009hf,Polchinski:2016hrw,Stoica:2018uli}. In particular, it suggests that information thrown into a black hole is not lost. The reason for this is that the AdS black hole undergoing evaporation is dual to unitary time evolution of a thermal state on the CFT side of the duality, which does not allow for information loss. The information thrown into a black hole is thus argued to be scrambled by some kind of unitary dynamics and then remitted via Hawking radiation \cite{Hayden:2007cs,Sekino:2008he,Lashkari:2011yi}. The question of how long one needs to wait for information thrown into a black hole to emerge in the subsequent Hawking radiation was first addressed in \cite{Hayden:2007cs}. It stated that information thrown into a black hole after the Page time would re-emerge within a scrambling time scale which is given by:
\begin{equation}
    t_{scr}\sim \beta\ln(S),
\end{equation}
where $\beta$ is the inverse Hawking temperature and $S$ is the number of degrees of freedom in the black hole which take part in scrambling. 

Usually in the context of AdS/CFT one considers black holes well beyond the Hawking-Page transition. These black holes, often referred to as large AdS black holes, are dual to large $N$ gauge theories \cite{Witten:1998zw,Maldacena:2001kr}. They have a horizon radius, $r_s$, that satisfies $r_s\gg L$ where $L$ is the AdS radius. A peculiar property of large AdS black holes is that they are thermally stable. This is due to the confining potential which comes from the asymptotics of AdS spacetimes. In such a case any Hawking radiation that the black hole emits reaches the conformal boundary and bounces back, being reabsorbed into the black hole. Eventually the black holes reaches stable equilibrium with the surrounding Hawking radiation and will not evaporate \cite{hawking1982,Hubeny:2009rc}. This makes large AdS black holes ill-suited to discuss the information paradox. To remedy this issue, it has been suggested to start with a large AdS black hole and then couple the bulk fields to an auxiliary field (called the evaporon) which carries energy away from the AdS black hole into an auxiliary system thereby allowing the black hole to evaporate \cite{Rocha:2008fe,Rocha:2009xy}.

In such constructions, it is the joint system of the reservoir and black hole which satisfy unitarity. Such constructions have been of recent interest in explorations of the information paradox. For example,  \cite{Penington:2019npb,Almheiri:2019psf} rely on such setups to show how information from the black hole gets released in the Hawking radiation (see \cite{Almheiri:2020cfm} for a recent overview of the literature). They use entanglement wedge reconstruction to show how information inside a black hole after the Page time scale is encoded in the subsequent Hawking radiation. In particular, Pennington showed that a small amount of information thrown into a black hole (after the Page time) will re-emerge in Hawking radiation after a time scale given by:
\begin{equation}
\label{temerge}
    t_{emerge}=\frac{\beta}{2\pi}\ln\left(\frac{2\pi C}{\beta \left|\frac{dr_s}{dt}\right|}\right),
\end{equation}
where $C$ can be thought of as the radial distance away from the horizon that one expects the Rindler description to hold, $dr_s/dt$ is the average rate of change of the horizon radius during evaporation, and $\beta$ is the inverse Hawking temperature. Moreover, as we shall review in Section \ref{PenningtonScrmabSec}, $t_{emerge}$ is the scrambling time scale discussed in \cite{Hayden:2007cs}. A key assumption that was made in the calculation was that radiation was being extracted close to the horizon by some type of ``super-observer'' in a non-local manner. Since the radiation was extracted sufficiently close to the horizon it was assumed that greybody factors can be ignored and the 2D Stefan-Boltzmann law was used for the evaporation rate:
\begin{equation}
    \frac{dM}{dt}=\frac{c_{evap} \pi}{12\beta^2},
\end{equation}
where $c_{evap}$ represents the number of modes being extracted near the horizon. Using this evaporation rate in conjunction with the first law of black hole thermodynamics ($dM=TdS$) gave an information emergence time of the form\footnote{We will review some of the details of the calculation in Section \ref{PenningtonScrmabSec}}:  
\begin{equation}
\label{HigherDimEnergenceTime}
    t_{emerge}\sim \frac{\beta}{2\pi}\ln\left( \frac{S-S_{ext}}{c_{evap}} \right).
\end{equation}
A similar result is also derived for 2D black holes in Jackiw-Teitelboim (JT) gravity studied in \cite{Almheiri:2019psf}. Which is given by:
\begin{equation}
\label{2DEmergenceTime}
    t_{emerge}\sim \frac{\beta}{2\pi} \ln\left(\frac{S-S_{ext}}{c}\right),
\end{equation}
where $c$ is the central charge (a measure of the degrees of freedom of a CFT) of a CFT that describes bulk matter in the 2D gravity theory. In light of the two results in Eqs. (\ref{HigherDimEnergenceTime} - \ref{2DEmergenceTime}) for the emergence time, it is tempting to make a rough identification of $c\sim c_{evap}$. The central charge, $c$, in Eq. (\ref{2DEmergenceTime}) seems to be a fixed parameter which does not appear to have any kind of dependence on quantities that characterize the black hole such as temperature. 

However, it is clear that in Pennington's setup $c_{evap}$ depends on details of where and how radiation is extracted near the horizon. For example, $c_{evap}$ should depend on how close one is extracting radiation near the horizon. The closer we are, the larger $c_{evap}$ can get. Furthermore, $c_{evap}$ will depend on the means by which one extracts radiation from the horizon; if we choose to place a surface at a radial distance $\delta r$ from the horizon with perfectly absorbing boundary conditions then $c_{evap}$ would be larger than if we chose some kind of semi-reflective boundary conditions. All these details will have some effect on the value of $c_{evap}$ and therefore on the evaporation rate. 

In light of these observations, we explore how the evaporation rate of a black hole depends on how close we extract radiation from the horizon. In this paper, we will model the ``super-observer'' using an absorptive screen placed close to the horizon. Roughly speaking, we assume that the screen can be understood from the prospective of the holographic renormalization group in AdS/CFT \cite{Freedman:1999gp,deBoer:2000cz}. At infinity we have a full UV complete (local) theory. The degrees of freedom on the screen and their dynamics are going to be viewed as a lower energy coarse grained version of the UV theory. We expect that the lower energy theory will become increasingly non-local as we push the screen closer to the horizon\footnote{We will evaluate this interpretation of the screen in more detail in Section \ref{NEC c-thm discussion} when we discuss the null energy condition for the screen and connections to the holographic c-theorem.}.   

To simplify considerations, we assume that the screen will absorb any radiation that reaches it \footnote{By doing this we are not actually defining the effective theory living on the screen that is consistent with some UV completed theory on the boundary. If we did make the effective theory on the screen consistent with a UV completed theory, we should not expect a perfectly absorptive screen. However, we still believe that a perfectly absorptive screen near the horizon is a reasonable approximation. In Section \ref{RigourousApproachtoGBF}, we propose a more rigorous way of defining how the screen should absorb radiation.}. In Section \ref{IntroGeneralizedGreyBodyFact}, we review how to calculate the average evaporation rate of a black hole and discuss how greybody factors affect this rate. By doing this we are able to clearly identify Pennington's $c_{evap}$ in terms of an infinite sum over angular momentum modes. We discuss how in two dimensions $c_{evap}$ in Eq. (\ref{HigherDimEnergenceTime}) can be reasonably identified with $c$ in Eq. (\ref{2DEmergenceTime}) with no further dependence on parameters that characterize the black hole. However, in higher dimensions we find that such a naive identification is not valid. We introduce the notion of a generalized greybody factor which quantifies the fraction of radiation that gets to a point at a radial distance $\delta r$ away from the horizon. At this distance away we introduce a perfectly absorbing screen which will absorb any radiation that hits it. We then write down an expression for the evaporation rate in terms of the generalized greybody factor. After doing this we restrict ourselves to massless scalar perturbations and write down a model for the generalized greybody factor which treats the effective potential as a ``hard wall.'' In Section \ref{AdSSchBHEvapRate}, we apply the hard wall model to AdS Schwarzschild black holes and find the evaporation rate. In Section \ref{NearExtEvapSec}, we discuss why the hard wall model for the generalized greybody factor is not sufficient for near extremal AdS Reissner–Nordstrom (RN) black holes. We motivate a correction that ``softens'' the wall and accounts for radiation being able to tunnel into the classically forbidden region. We then provide an estimate using this modified model for the evaporation rate of near extremal AdS RN black holes. In Section \ref{PenningtonScrmabSec}, we review Pennington's calculation of $t_{emerge}$ and then use the modified evaporation rates that we calculated in Sec. \ref{Sec2EvapRates} and find $t_{emerge}$. In particular, for AdS black holes with $r_s/L\ll 1$ we find results that agree with Pennington's calculation up to some logarithmic correction which depends on how far we choose to extract radiation. However, in the case of $r_s/L\gg 1$ we find a slightly different result; the argument that goes into the Log is not the entropy of the entire horizon, but rather the entropy of a cell of size $L$ controlled by the AdS radius (in addition to the usual logarithmic correction which depends on the extraction radius). In Section \ref{InfoEmAsScr}, we discuss the subtleties involved in choosing the $\beta$ dependence of the subleading Log correction for near extremal black holes. By fixing the proper distance between the screen and horizon we find that $t_{emerge}$ is consistent with the scrambling time for near extremal black holes (up to a sub-leading Log correction that has no further dependence on the temperature of the black hole). We speculate that fixing the proper radial distance of the screen from the horizon to corresponds to fixing the energy scale of the effective holographic theory on the screen. In Section {\ref{RigourousApproachtoGBF}} we formulate a more rigorous framework to calculate how the screen will absorb Hawking radiation. This is done by viewing the screen as an interface which patches the interior black hole spacetime to an exterior ``reservoir'' spacetime. By doing this we reduce the problem of finding how the screen absorbs the radiation to a calculation of finding the transmission amplitude of scalar perturbations through an effective potential. We argue that by using this approach one should recover the results in reasonable agreement with the toy models discussed in this paper. In Section \ref{NEC c-thm discussion} we briefly review the holographic $c$-function and the role that the null energy condition (NEC) plays in its formulation. We then consider the radial NEC for the matter that makes up the screen and show that it satisfies the radial NEC a finite distance from the horizon as long as the AdS radius of the spacetime enclosed by the screen is smaller than the AdS radius of the exterior spacetime. This provides a heuristic way to quantify the effective coarse-grained degrees of freedom as the screen is moved toward the horizon. In Section \ref{MiningSection} we discuss how extracting Hawking radiation near the horizon of an AdS black hole can be tied in with discussions of black hole mining. We show that the radial NEC places non-trivial constraints on how close the screen is allowed to be to the horizon. The constraints show that small AdS black holes cannot be mined by placing a screen very close to the horizon. However, mining for very large AdS black holes is possible since the screen can be placed very close to the horizon without violations of the radial NEC. We compute how long it takes for a very large AdS black hole to transition to a small AdS black hole through screen mining. We estimate that to leading order a the transition time (in units of the AdS radius) is given by the Bekenstein-Hawking entropy of an AdS radius sized cell.

We then conclude this work by summarizing the major results of this paper as well as some outstanding questions and issues which can be explored further.

\section{Changing Evaporation Rates via Near Horizon Extraction}
\label{Sec2EvapRates}
\subsection{Modelling Hawking Radiation Extraction Through Generalized Greybody Factors}
\label{IntroGeneralizedGreyBodyFact}

It is well known that close to the horizon, a black hole will emit radiation as a black body. However, by the time this radiation reaches an observer very far away from the black hole the spectrum of the radiation is modified. This is because the black hole generates a non-trivial potential that perturbations travelling through the background will experience, resulting in partial reflection and transmission of perturbations. These effects are contained in greybody factors and they have a non-trivial effect on the evaporation rate of a black hole. We will review the basics of how greybody factors affect the the evaporation rate. We will then introduce the notion of a generalized greybody factor which will depend on how far one is extracting radiation from the horizon.  

We begin with the well known result which describes the occupation number distribution of Hawking quanta emitted by a black hole (not accounting for greybody factors):
\begin{equation}
    \braket{n(\omega)}_{\pm}=\frac{1}{e^{\beta\omega}\pm 1}.
\end{equation}
The plus is for fermionic Hawking quanta and the minus is for bosonic Hawking quanta. For the sake of simplicity we will restrict ourselves to bosonic quanta in this paper. The total evaporation rate (ignoring greybody factors) of the black hole is given by:
\begin{equation}
    \frac{dM}{dt}= \frac{1}{2\pi}\sum_{\ell}N_\ell\int_0^{\infty}N_b\omega\braket{n(\omega)}_-d\omega =\frac{1}{2\pi}\sum_{\ell} N_\ell\int^{\infty}_{0}\frac{N_b\omega}{e^{\beta \omega}-1}d\omega=\frac{N_b\pi}{12 \beta^2}\sum_\ell N_\ell,
\end{equation}
where $N_b$ is the number of different bosonic species and $N_\ell$ is the degeneracy of the $\ell$-th hyper-spherical harmonic\footnote{To understand why $N_\ell$ is present recall that the solution to the massless scalar wave equation in a spherically symmetric background can be decomposed as a product $\Psi(t,r,\vec{\phi})=e^{-i\omega t}r^{(1-d)/2}\psi(r)\Phi_{\ell}(\vec{\phi})$ where $\Phi_\ell$ are hyper-spherical harmonics for a given $\ell$ angular momentum mode there are $N_\ell$ degenerate eigenfunctions. In particular, we identify Pennington's $c_{evap}= N_b\sum_{\ell} N_{\ell}$.}. Note that we recover the 2D Stefan-Boltzmann law used by Pennington with the identification, $c_{evap}= N_b\sum_{\ell} N_{\ell}$. This is only finite in 2D where the sum over $\ell$ disappears and we are left with $c_{evap}=N_b$ which does not depend on the parameters that characterize the black hole (or even the exact position of the screen) this is similar to the behaviour of $c$ in Eq. (\ref{2DEmergenceTime}) which we discussed in Section \ref{introsec}. 

In higher dimensions the sum persists and will be divergent resulting in an infinite evaporation rate. The effective potential near the horizon is essential for understanding how the divergence is regulated in higher dimensions. Generally speaking, if we extract Hawking radiation a finite radial distance $\delta r$ from the horizon we should expect some fraction of the total radiation emitted by the black hole to to reach $r=r_s+\delta r$. This is due to the fact that the effective potential is only zero at the horizon and strictly increases (at least in some neighborhood of the horizon). The larger $\ell$ is the more quickly it increases, this causes the higher angular momentum modes to reflect back into the black hole, effectively placing a cutoff over the sum of angular momentum modes which result in a finite evaporation rate.

We define the generalized greybody factor, $\gamma_\ell(\omega,\delta r)$, for each $\ell$. It quantifies the fraction of radiation that gets to some surface a finite distance $\delta r$ from the horizon\footnote{In particular $\lim_{\delta r\to \infty}\gamma_{\ell}(\omega,\delta r$) will reproduce the greybody factors that are usually discussed in the context of an observer sitting at asymptotic infinity.}. If the absorptive surface is sitting at $r=r_s+\delta r$ then, the generalized greybody factor represents the fraction of energy absorbed by the screen from the $\ell$-th mode. Then the total rate at which the black hole losses mass is given by:
\begin{equation}
\label{GenBHEvapRate}
    \frac{dM}{dt}= \frac{1}{2\pi}\sum_\ell\int^{\infty}_0\frac{N_\ell N_b\gamma_\ell(\omega,\delta r)\omega}{e^{\beta\omega}-1}d\omega.
\end{equation}
The generalized greybody factor will be essential in regulating the infinite sum over $\ell$. In general, we can compute $\gamma_\ell(\omega,\delta r)$ by considering the wave equation on the black hole background. However, doing this analytically is difficult. To circumvent this issue we will introduce models for the generalized greybody factor which will capture the essential physics of the situation near the horizon. 

\begin{figure}[h!]
\centering
\includegraphics[width=150mm]{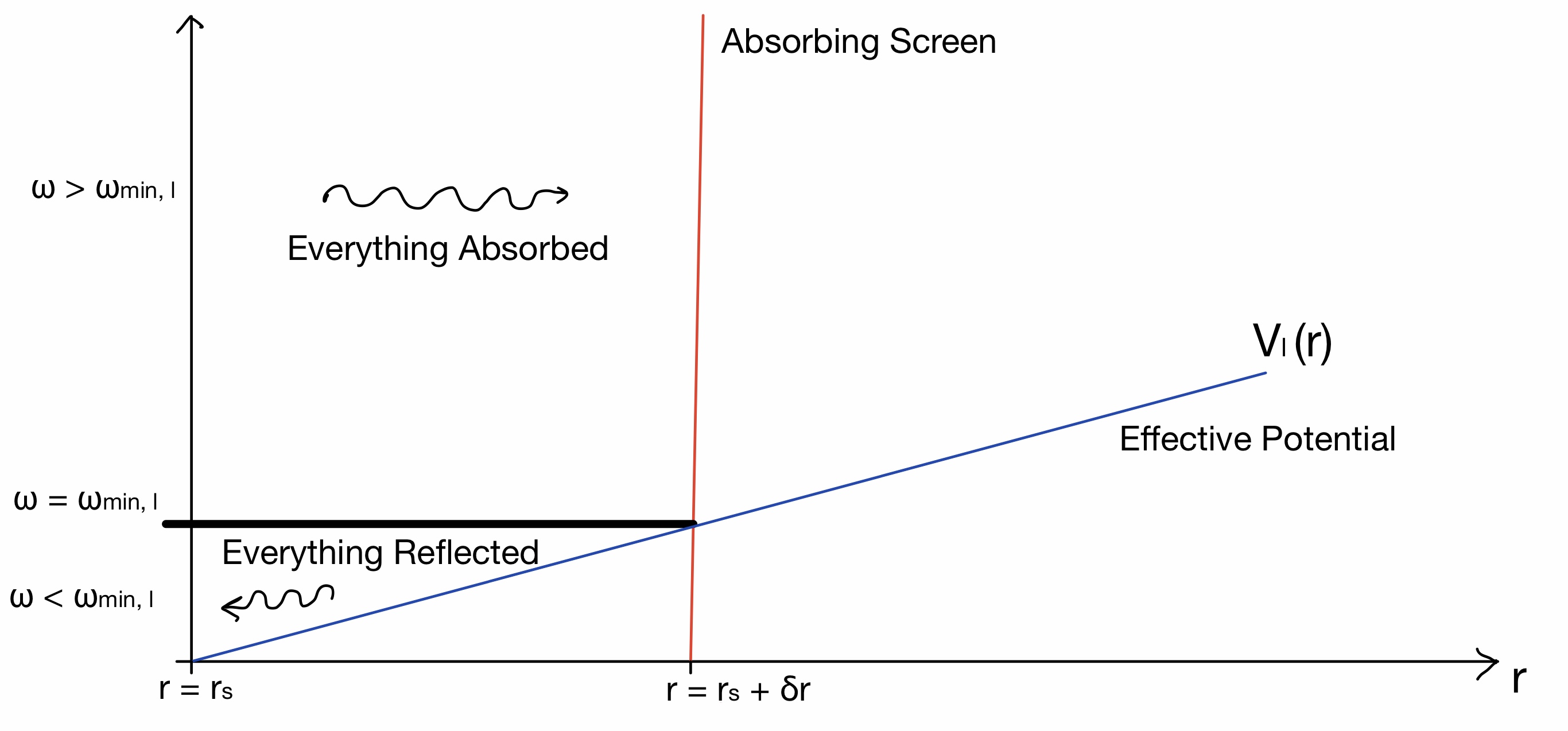}
\caption{Above is a depiction of how perturbations behave near the horizon with a generalized greybody factor given in Eq. (\ref{STEPGreyBodyFact}). Near the horizon the Potential $V_{\ell}(r)$ is linear and is depicted by the solid blue line. The slope of the blue line increases with $\ell$. The absorptive boundary is depicted by the vertical red line at $r=r_s+\delta r$. The thick black line is a lower bound for the frequency of radiation that gets absorbed. Everything below the thick line has frequency $\omega<\omega_{min,\ell}=\sqrt{V_{\ell}(r_s+\delta r)}$ and cannot get to the absorbing surface, it bounces off the potential and gets reabsorbed. Everything above the thick line has frequency $\omega>\omega_{min,\ell}$ and is able to reach the absorptive surface and gets completely absorbed.        \label{HardWallFig} }
\end{figure}

For the sake of concreteness we will consider the massless scalar wave equation for a spherically symmetric black hole background in $d+1$-dimensions \footnote{These black hole spacetimes will generally have a metric of the following form $ds^2=-f(r)dt^2+dr^2/f(r)+r^2 d\Omega_{d-1}^2$.}.
We are interested in the radial part of the solution which can be shown to obey the Regge-Wheeler equation:
\begin{equation}
    \frac{d^2\psi}{dr_*^2}+(\omega^2-V_\ell)\mathcal{\psi}=0,
\end{equation}
where $r_*$ is the tortoise coordinate defined by the relation $dr_*=\frac{dr}{f(r)}$, and $V_\ell$ is the effective potential given by:
\begin{equation}
    V_\ell(r)=f(r)\left[\frac{d-1}{2r}\frac{df}{dr}+\frac{(d-1)(d-3)}{4r^2}f(r)+\frac{\ell(\ell+d-2)}{r^2}\right].
\end{equation}
If we choose to extract radiation close to the horizon (i.e. $r-r_s\ll r_s$) we can approximate $V_\ell$ to linear order as:
\begin{equation}
   V_\ell(r)\simeq  \frac{4\pi}{\beta}\left[ \frac{(d-1)2\pi}{\beta r_s}+\frac{\ell(\ell+d-2)}{r_s^2} \right](r-r_s)+...,
\end{equation}
where $\beta$ is the inverse Hawking temperature. We place a perfectly absorbing surface at $r-r_s=\delta r$ where $\delta r/r_s\ll 1$. Now consider the quantity:
\begin{equation}
\label{OmegaminusV}
    \omega^2-V_\ell(r)\simeq \omega^2-\frac{4\pi}{\beta}\left[ \frac{(d-1)2\pi}{\beta r_s}+\frac{\ell(\ell+d-2)}{r_s^2} \right](r-r_s)+...
\end{equation}
As long as $\omega^2\gg V_\ell$ we should be able to ignore the effects of $V_\ell$; the radiation will experience little to no hindrance to get to the absorbing screen we place near the horizon (i.e. $\gamma_\ell(\omega,\delta r)\sim 1$). However, once $\omega^2\leq V_\ell$ we should expect most of the radiation to be reflected back into the black hole and reabsorbed (i.e. $\gamma_\ell(\omega,\delta r)\sim 0$). We depict the scenario in Figure. \ref{HardWallFig}.
We model this sort of ``hard wall'' potential by introducing the following generalized greybody factor:
\begin{equation}
\label{STEPGreyBodyFact}
    \gamma_\ell(\omega,\delta r)=\Theta\left[\omega^2-V_\ell(r_s+\delta r)\right],
\end{equation}
where $\Theta$ is the Heaviside step function. Using this model the evaporation rate using Eq. (\ref{GenBHEvapRate}) is given by:
\begin{equation}
\label{EvapRateNearHorizon}
    \frac{dM}{dt}=\frac{ N_b}{2\pi}\sum_{\ell=0}^\infty \int_{\omega_{min,\ell}}^\infty \frac{N_\ell\omega}{e^{\beta\omega}-1}d\omega=\frac{N_b}{2\pi \beta^2}\sum_{\ell=0}^{\infty}\left(N_\ell\int_{x_{min,\ell}}^{\infty}\frac{x}{e^{x}-1}dx\right),
\end{equation}
where $\omega_{min,\ell}$ satisfies:
\begin{equation}
    \omega_{min,\ell}^2-V_\ell(r_s+\delta r)=0.
\end{equation}
In the next section, we use this model to find the evaporation rate of AdS Schwarzschild black holes. We will also use a similar model with some adjustments to calculate the evaporation rate of near extremal AdS RN black holes. 
\subsection{AdS Schwarzschild Black Holes}
\label{AdSSchBHEvapRate}
\begin{figure}[h!]
\centering
\includegraphics[width=150mm]{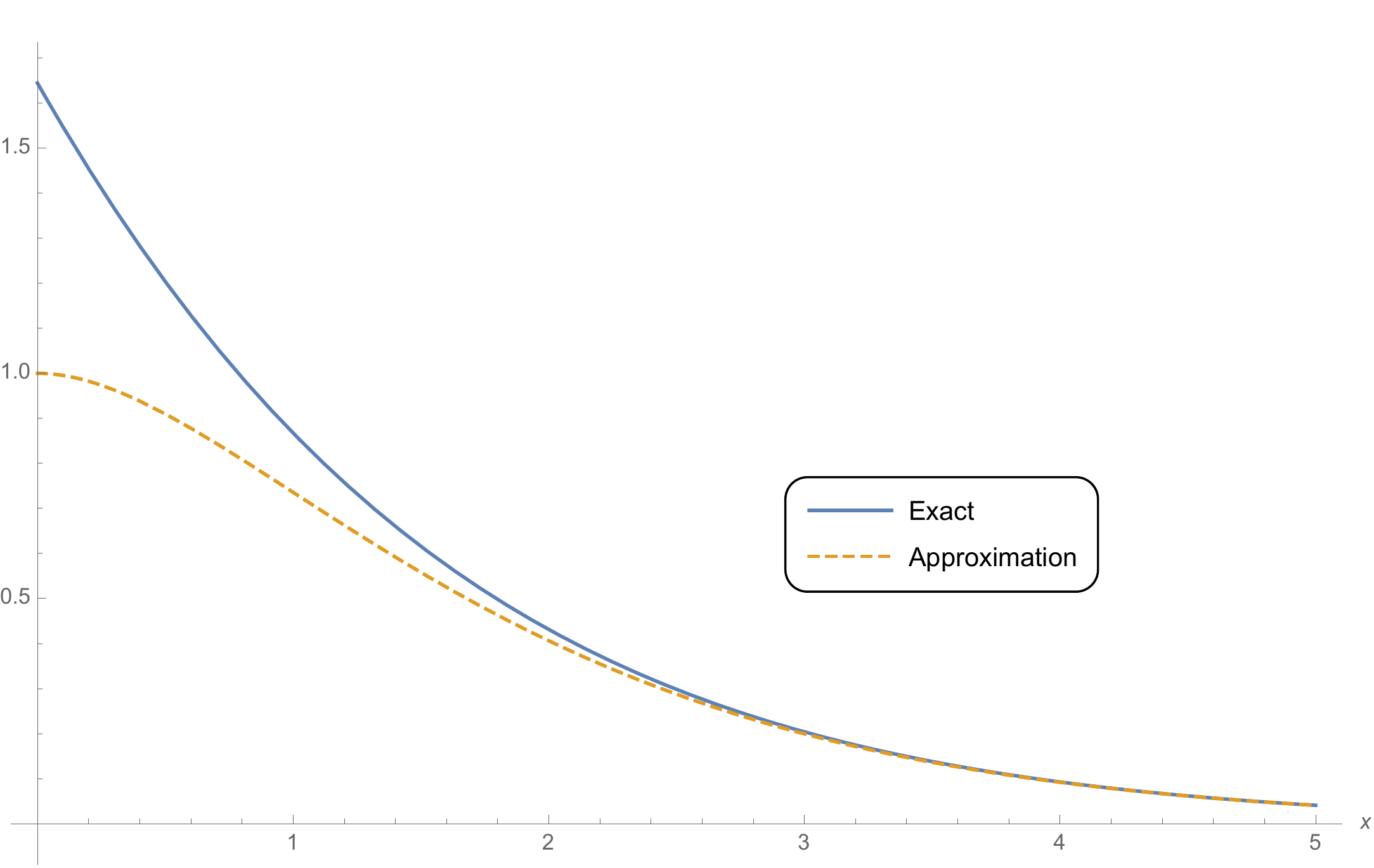}
\caption{The solid blue line labeled ``Exact'' is the left hand side of Eq. (\ref{approxExact}) and the dotted yellow line labeled ``Approximation'' is the right hand side of Eq. (\ref{approxExact}).   \label{ApproxExactFig} }
\end{figure}

In this section we will estimate the evaporation rate of a $d+1$ - dimensional AdS Schwarzschild black hole\footnote{The AdS Schwarzschild black hole has the following line element $ds^2=-f(r)dt^2+\frac{dr^2}{f(r)}+r^2d\Omega^2_{d-1}$, where $f(r)=1+\frac{r^2}{L^2}-\left(\frac{r_s}{r}\right)^{d-2}\left(1+\frac{r_s^2}{L^2}\right)$. The Hawking temperature of these black holes are given by $T_H=\frac{dr_s^2+(d-2)L^2}{4\pi r_s L^2}$}. We start by doing the integral in Eq. (\ref{EvapRateNearHorizon}) and obtain the following result: 
\begin{equation}
    \frac{dM}{dt}=\frac{N_b}{2\pi\beta^2}\sum_{\ell=0}^{\infty}N_\ell\left[Li_2\left(e^{-x_{min,\ell}}\right)-x_{min,\ell}\ln\left(1-e^{-x_{min,\ell}}\right)\right],
\end{equation}
where the $x_{min,\ell}$ is given by:
\begin{equation}
    x_{min,\ell}=\beta \omega_{min,\ell}=\sqrt{4\pi\left( 2\pi(d-1)+\frac{\beta \ell(\ell+d-2)}{r_s} \right)\frac{\delta r}{r_s}},
\end{equation}
and $Li_m(x)$ is the $m$-th order polylog function in $x$. We estimate the value of the series as follows. We note that $x_{min,\ell}$ increases with $\ell$. So for sufficiently large $\ell$ we have the following leading order approximation for the evaporation rate:
\begin{equation}
\label{approxExact}
    Li_2\left( e^{-x_{min,\ell}} \right)-x_{min,\ell}\ln\left(1-e^{-x_{min,\ell}}\right)\simeq (1+x_{min,\ell})e^{-x_{min,\ell}}+\mathcal{O}\left(e^{-2x_{min,\ell}}\right).
\end{equation}
We expect the approximation used above is accurate for very large values of $\ell$. In Figure \ref{ApproxExactFig} we plot the exact function and the approximation.
If we use the approximated function for any $\ell \geq 1$ we expect to get a reasonable estimate for the series (accurate within an order of magnitude). We approximate the degeneracy of angular momentum modes as $N_{\ell}\sim \ell^{d-2}$ so we have:
\begin{equation}
\begin{split}
   & \frac{dM}{dt}\approx \frac{N_b}{2\pi\beta^2}\left[\frac{\pi^2}{6}+\sum_{\ell=1}^{\infty}\left[ \ell^{d-2}\left(1+x_{min,\ell}\right) e^{-x_{min,\ell}}\right]  \right],\\
    \end{split}
\end{equation}
where the $\pi^2/6$ term comes from the $\ell=0$ mode in the limit where $\delta r/r_s\to 0$. To do the sum over modes with $\ell\geq 1$ in closed form we need to make an additional approximation that simplifies the functional form of $x_{min,\ell}$:
\begin{equation}
\begin{split}
&x_{min,\ell}\approx \alpha\ell\\
&\alpha=\sqrt{\frac{4\pi \beta \delta r}{r_s^2}}\ll 1.\\
\end{split}
\end{equation}
This approximation comes from the leading order expansion of $x_{min,\ell}$ for large $\ell$. We can then do the sum in closed form and get:
\begin{equation}
\begin{split}
   & \frac{dM}{dt}\approx \frac{N_b}{2\pi\beta^2}\left[\frac{\pi^2}{6}+\sum_{\ell=1}^\infty\ell^{d-2}\left(1+\alpha \ell \right)e^{-\alpha \ell} \right]\\
   &=\frac{N_b}{2\pi\beta^2}\left[\frac{\pi^2}{6}+\alpha Li_{1-d}\left( e^{-\alpha} \right)+Li_{2-d}\left(e^{-\alpha}\right) \right]. \\
    \end{split}
\end{equation}
If $\delta r/r_s$ is sufficiently small (i.e. the screen is sufficiently close to the horizon) we can do a series expansion in $\alpha$ near zero. The leading order contribution to the estimate for the evaporation rate is given by:
\begin{equation}
\label{fullevaprateAdSSch}
    \frac{dM}{dt}\approx \frac{N_b}{2\pi\beta^2}\left[ d(d-2)!\left(\frac{r_s^2}{4\pi\beta \delta r}\right)^{\frac{d-1}{2}} +\mathcal{O}(1)\right].
\end{equation}
In Appendix \ref{EvapApproxAppendix} we do a detailed comparison of our leading order estimate for the evaporation rate given in Eq. (\ref{fullevaprateAdSSch}) with numerical calculations. We find that our estimate for the series agrees with numerical results up to a pre-factor of order 1 (see Tables \ref{Tabled3delta10m4} - \ref{Tabled10delta10m4}). Notice that the $\ell=0$ mode is a order one correction to the leading order term which is much larger if $\delta r/r_s$ is sufficiently small (i.e. the screen is placed sufficiently close). To avoid clutter in our leading order expression we will define a dimension dependent coefficient $\mathcal{A}_d$ and write the evaporation rate as:

\begin{equation}
\label{SchAdSBHevaprate}
\begin{split}
    &\frac{dM}{dt}\approx \frac{ \mathcal{A}_d N_b}{\beta^2}\left(\frac{r_s^2}{\beta \delta r}\right)^{\frac{d-1}{2}}\\
    &\mathcal{A}_d=\frac{d(d-2)!\left(4\pi \right)^{\frac{1-d}{2}}}{2\pi}. \\
    \end{split}
\end{equation}

\subsection{Near Extremal AdS RN Black Holes}
\label{NearExtEvapSec}
Now lets consider $d+1$ - dimensional  near extremal AdS RN black holes\footnote{The $d+1$ - dimensional AdS RN black hole has a line element of the form $ds^2=-f(r)dt^2+\frac{dr^2}{f(r)}+r^2d\Omega_{d-1}^2$, where $f(r)=\left(1-\frac{r_s^{d-2}}{r^{d-2}}\right)\left(1-\frac{Q^2}{r^{d-2}r_s^{d-2}}\right)+\frac{r^2}{L^2}\left(1-\frac{r_s^d}{r^d}\right)$, where $r_s$ is the radius of the horizon and $Q$ is the charge of the black hole. The black hole is extremal when the charge and horizon radius satisfy the following relation, $Q^2=r_s^{2(d-2)}\left(1+\frac{d}{d-2}\frac{r_s^2}{L^2}\right)$, this occurs when $T_H=\frac{f'(r_s)}{4\pi}=0$. }. We analyze how the evaporation rate depends on where we extract radiation near the horizon. In this case we should expand $V_\ell$ to second order. This is because the first order expansion of $V_{\ell}$ is proportional the temperature which will go to zero in the extremal limit. Sufficiently close to the extremal regime the second order term will dictate the leading order behaviour of the potential close to the horizon.
 \begin{equation}
 \begin{split}
  & V_\ell(r)\simeq  V_1(r-r_s)+\frac{V_2}{2}(r-r_s)^2+...\\
  &V_1= \frac{4\pi}{\beta}\left[ \frac{(d-1)2\pi}{\beta r_s}+\frac{\ell(\ell+d-2)}{r_s^2} \right]\\
  &V_2=f''(r_s)\frac{\ell(\ell+d-2)}{r_s^2}\\
  &+\frac{4\pi}{\beta}\left[2\frac{d}{dr}\left( \frac{d-1}{2r}\frac{df}{dr}+\frac{(d-1)(d-3)f(r)}{4r^2}+\frac{\ell(\ell+d-2)}{r^2} \right) +f''(r)\frac{d-1}{2r}  \right]\bigg\vert_{r=r_s}. \\
\end{split}
\end{equation}
The expansion above will be valid if $r-r_s\ll r_s$. Sufficiently close to the extremal regime we will have the leading order contribution equal to:
\begin{equation}
\begin{split}
    &V_\ell(r)= \frac{f''_{ext}(r_s)}{2}\frac{\ell(\ell+d-2)}{r_s^2}(r-r_s)^2. \\
    \end{split}
\end{equation}
As before, we can consider placing a perfectly absorbing surface a radial distance $\delta r$ from the horizon. If we decide to use the Heaviside step model in Eq. (\ref{STEPGreyBodyFact}) then we will need to do the integral in Eq. (\ref{EvapRateNearHorizon}) with the lower bound:
\begin{equation}
\begin{split}
  &x_{min,\ell}= \beta\omega_{\min,\ell}= \frac{\beta \delta r}{r_s}\sqrt{\frac{f''_{ext}(r_s)\ell(\ell+d-2)}{2}}. \\
  \end{split}
\end{equation}

\begin{figure}[h!]
\centering
\includegraphics[width=150mm]{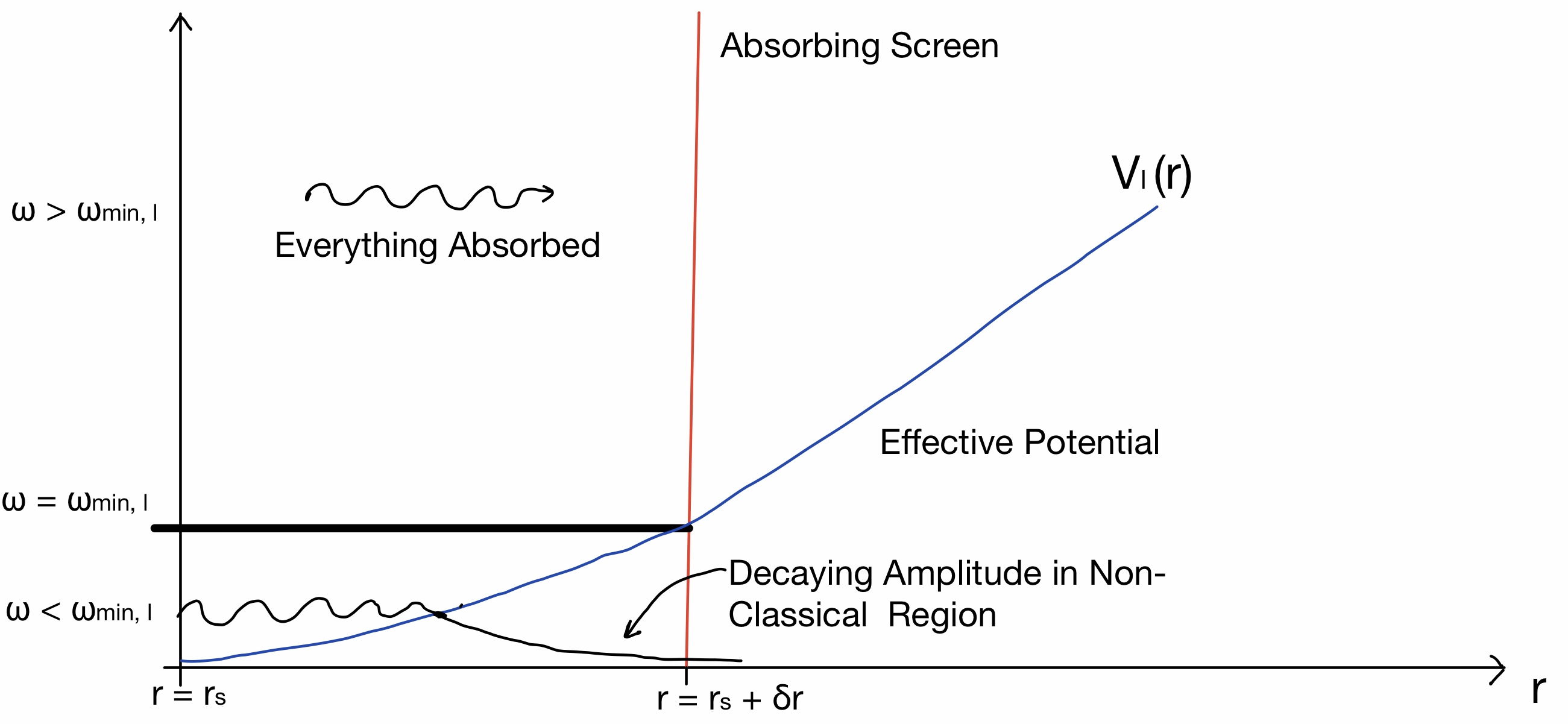}
\caption{Above is a depiction of how perturbations behave near the horizon with a generalized greybody factor given in Eq. (\ref{GeneralizedGBFNearExt}). For $\omega_{min,\ell}$ the model is unchanged and everything is absorbed. However, for $\omega<\omega_{min,\ell}$ we account for the wave-like behaviour of the solution which allows for the solution to tunnel into the classically forbidden region. The amplitude the the solution would decay as some power law after the classical turning point. We estimate the amount of energy that tunnels to the absorptive surface by taking the ratio between the amplitude of the solution at the turning point and the amplitude at the absorptive surface. Doing this gives a power law suppression of the generalized greybody factor for $\omega<\omega_{min,\ell}$ in Eq. (\ref{GeneralizedGBFNearExt}).       \label{SoftWallFig} }
\end{figure}

Unlike the non-extremal case we discussed previously the lower bound is much larger than unity sufficiently close to the extremal regime. This means that we are well into the exponentially decaying tail of the integrand. Recall that the Heaviside step function model was used to simulate the effective potential as a ``hard'' wall. In reality we know that the waves can actually enter the classically forbidden region. The amplitude of the solution will decay through some power law in the classically forbidden region. By the time a wave with $\omega<\omega_{min,\ell}$ reaches the absorptive surface its amplitude would be power law suppressed as depicted in Fig \ref{SoftWallFig}. 
The Heaviside model completely disregards these effects. This would be okay if the contribution of modes with $\omega\geq \omega_{min,\ell}$ was not exponentially suppressed, but since it is suppressed in the near extremal regime we need to consider the effects of $\omega<\omega_{min,\ell}$. Therefore, for a near extremal black hole we need a generalized greybody factor of the form:
\begin{equation}
    \gamma_{\ell}(\omega,\delta r)=\Theta(\omega-\omega_{min,\ell})+\left( \frac{\omega}{\omega_{min,\ell}} \right)^{q(\ell)}\Theta\left( \omega_{min,\ell}-\omega \right),
\end{equation}
where $q(\ell)$ is some function of $\ell$ which will be determined by analyzing the dynamics of the perturbations near the horizon and gives us the power law decay we need. The details of how to obtain a reasonable model for $q(\ell)$ for scalar wave perturbations is detailed in Appendix \ref{PowerLawAppendix}. The result for $\ell\geq 1$ is:
\begin{equation}
\begin{split}
\label{GeneralizedGBFNearExt}
    &\gamma_{\ell}(\omega,\delta r)=\Theta(\omega-\omega_{min,\ell})+\left( \frac{\omega}{\omega_{min,\ell}} \right)^{2\nu_\ell+1}\Theta\left( \omega_{min,\ell}-\omega \right)\\
    &\nu_{\ell}=\sqrt{\frac{1}{4}+\alpha_{\ell}^2}\\
    &\alpha_{\ell}^2=\frac{2\ell(\ell+d-2)}{r_s^2f''_{ext}(r_s)}, \\
\end{split} 
\end{equation}
where $f''_{ext}(r_s)$ is the second derivative of $f(r)$ evaluated at the horizon radius $r_s$, in the limit where the Hawking temperature goes to zero. Using this, the expression for the contribution to the evaporation rate for $\ell\geq 1$ is given by\footnote{We are not including the $\ell=0$ mode in this section, in Appendix \ref{elleq0mode} we treat the $\ell=0$ mode. We show that for very large AdS black holes sufficiently close to the extremal regime the $\ell=0$ contribution is sub-leading compared to the contribution of modes with $\ell\geq 1$. For very small AdS black holes and asymptotically flat black holes the $\ell=0$ mode has a leading order effect on the evaporation rate when $d\leq 8$ in the near extremal regime, we discuss this point and also discuss its consequences on the information re-emergence time. Overall we find that including the $\ell=0$  results in minor changes in the expressions for information re-emergence time which are consistent with the usual scrambling time scales for nearly extremal AdS RN black holes.}:
\begin{equation}
\label{NearExtEvapRateInteg}
\begin{split}
\frac{dM}{dt}&= \frac{N_b}{2\pi} \sum_{\ell=1}^\infty  \left[\int_0^{\omega_{min,\ell}} \frac{N_{\ell} \omega }{e^{\beta\omega}-1}\left(\frac{\omega}{\omega_{min,\ell}}\right)^{2\nu_\ell+1}d\omega +\int_{\omega_{\min,\ell}}^\infty \frac{N_{\ell}\omega}{e^{\beta\omega}-1}d\omega \right]\\
&=\frac{N_b}{2\pi}\sum_{\ell=1}^\infty \omega_{min,\ell}^2 N_\ell \left[ \int_0^1\frac{ \epsilon^{2(\nu_\ell+1)}}{e^{\beta \omega_{min,\ell}\epsilon}-1} d\epsilon +\int_{1}^\infty \frac{\epsilon}{e^{\beta\omega_{min,\ell}\epsilon}-1}d\epsilon  \right]. \\
\end{split}
\end{equation}
With some work detailed in Appendix \ref{IntegralsNearExt} we can do the integrals in closed form and write the total evaporation rate as the following series over $\ell$:

\begin{equation}
\begin{split}
\label{SeriesRepofEvapRateNearExt}
   &\frac{dM}{dt}=\frac{N_b}{2\pi\beta^2}\sum_{\ell=1}^{\infty}\left[ S_{tun,\ell}+S_{ntun,\ell} \right]\\
   &S_{tun,\ell}= N_\ell \left[\frac{\Gamma\left( 3+\sqrt{1+4\alpha_{\ell}^2} \right)Li_{3+\sqrt{1+4\alpha_\ell^2}}\left(1\right)}{x_{min,\ell}^{1+\sqrt{1+4\alpha_\ell^2}}} \right] \\
   &S_{ntun,\ell}= N_\ell\left[Li_2\left(e^{-x_{min,\ell}}\right)-x_{min,\ell}\ln\left(1-e^{-x_{min,\ell}}\right)\right]\\
   &x_{min,\ell}=\beta\omega_{min,\ell}=\frac{\beta\delta r f''_{ext}(r_s)\alpha_{\ell}}{2}=\frac{\beta\delta r}{r_s}\sqrt{\frac{f''_{ext}(r_s)\ell(\ell+d-2)}{2}}. \\
   \end{split}
\end{equation}
The terms $S_{tun,\ell}$ in Eq. (\ref{SeriesRepofEvapRateNearExt}) represents the contribution of to the evaporation rate of modes that tunnel through the effective potential. The terms $S_{ntun,\ell}$ in Eq. (\ref{SeriesRepofEvapRateNearExt}) represent the contribution to the evaporation rate of modes that do not need to tunnel through the barrier to reach the screen. When we are sufficiently close to the extremal regime we can show that $S_{tun,\ell}\gg S_{ntun,\ell}$ (this point is discussed in Appendix \ref{TunVSnonTun}). This means that sufficiently close to the extremal regime we can ignore $S_{ntun,\ell}$ and write:
\begin{equation}
\label{OnlyTunEvaprateSer}
   \frac{dM}{dt}\approx \frac{N_b}{2\pi\beta^2}\sum_{\ell=1}^\infty S_{tun,\ell}=\frac{N_b}{2\pi\beta^2}\sum_{\ell=1}^\infty \ell^{d-2} \left[\frac{\Gamma\left( 3+\sqrt{1+4\alpha_{\ell}^2} \right)Li_{3+\sqrt{1+4\alpha_\ell^2}}\left(1\right)}{x_{min,\ell}^{1+\sqrt{1+4\alpha_\ell^2}}} \right],
\end{equation}
where we used $N_\ell\sim \ell^{d-2}$. We cannot evaluate the series in closed form so we resort to additional approximations. 

We begin by considering the case of very large AdS black holes where $r_s/L\gg 1$. In this case we have:

\begin{equation}
    \alpha_{\ell+1}-\alpha_{\ell}= \frac{\sqrt{2\ell(\ell+d-2)}-\sqrt{2(\ell+1)(\ell+d-1)}}{\sqrt{d(d-1)}}\frac{L}{r_s}\sim \left(\frac{r_s}{L}\right)^{-1}\ll 1.
\end{equation}
The spacing between consecutive $\alpha_{\ell}$ becomes smaller as the AdS black holes we are considering become larger relative to the AdS radius. In this case, we estimate the sum using an integral as follows: 
\begin{equation}
\label{integrallargeextBH}
\begin{split}
   &\frac{dM}{dt}\approx \frac{N_b}{2\pi\beta^2}\int_{1}^{\infty} \ell^{d-2}\left[\frac{\Gamma\left( 3+\sqrt{1+4\alpha_{\ell}^2} \right)Li_{3+\sqrt{1+4\alpha_\ell^2}}\left(1\right)}{x_{min,\ell}^{1+\sqrt{1+4\alpha_\ell^2}}} \right]d\ell\\
   &=\frac{N_b}{2\pi\beta^2}\int_{\alpha_1}^{\infty} \left( \frac{d-2}{2}\left[-1+\sqrt{1+\frac{2r_s^2f_{ext}''(r_s)\alpha_{\ell}^2}{(d-2)^2}}\right] \right)^{d-2} \\
    &\times \left[ \frac{\Gamma\left(3+\sqrt{1+4\alpha_{\ell}^2}\right)Li_{3+\sqrt{1+4\alpha_{\ell}^2}}(1)}{\left(\frac{\beta\delta r f''_{ext}(r_s)\alpha_\ell}{2}\right)^{1+\sqrt{1+4\alpha_\ell^2}}} \right] \left(\frac{\alpha_{\ell}r_s^2f''_{ext}(r_s)}{(d-2)\sqrt{1+\frac{2r_s^2f''_{ext}(r_s)\alpha_{\ell}^2}{(d-2)^2}}}\right)d\alpha_\ell, \\
   \end{split}
\end{equation}
where in the last line we simply changed the variables of integration from $\ell$ to $\alpha_{\ell}$ using the definition of $\alpha_{\ell}$ in Eq. (\ref{GeneralizedGBFNearExt}). The lower bound of integration is $\alpha_1$, equal to:
\begin{equation}
\alpha_1=\sqrt{\frac{d-1}{(d-2)^2+d(d-1)\frac{r_s^2}{L^2}}}\approx \sqrt{\frac{L^2}{dr_s^2}}\ll 1.  
\end{equation}
For very large AdS black holes ($d\geq 4$) we use the following approximation for the integrand: (The steps to arrive at this approximation are described in Appendix \ref{NumericalAnalysisNearExtrBH} we also make plots to show that the the approximation will become more accurate as $r_s/L$ becomes larger.):
\begin{equation}
\label{evapRateNearExtLargeAdSBHdgeq4}
\begin{split}
    &\frac{dM}{dt}\bigg\vert_{d\geq 4}\approx\frac{N_b}{2\pi\beta^2} \int_{0}^\infty\frac{\pi^4}{15}\left(\frac{r_s^2f''_{ext}(r_s)}{2}\right)^{\frac{d-1}{2}}\left(\frac{\beta\delta r f''_{ext}(r_s)}{2}\right)^{-2-2\alpha_\ell^2}\alpha_\ell^{d-4}d\alpha_{\ell}\\
    &=\frac{N_b}{2\pi\beta^2}\frac{\pi^4}{15}\left(\frac{r_s^2f''_{ext}(r_s)}{4}\right)^{\frac{d-1}{2}}\left[\ln\left(\frac{\beta\delta rf''_{ext}(r_s)}{2}\right)\right]^{\frac{3-d}{2}}\left(\frac{2}{\beta\delta r f''_{ext}(r_s)}\right)^2 \Gamma\left(\frac{d-3}{2}\right)\\
    &\approx\frac{N_b}{2\pi\beta^2} \frac{\pi^4}{15}\left(\frac{2d(d-1)r_s^2}{4L^2}\right)^{\frac{d-1}{2}}\left[\ln\left(\frac{d(d-1)\beta\delta r}{L^2}\right)\right]^{\frac{3-d}{2}}\left(\frac{L^2}{d(d-1)\beta\delta r}\right)^2\Gamma\left(\frac{d-3}{2}\right). \\
    \end{split}
\end{equation}
Note that the expression above is ill defined for $d=3$. This is because we approximated the lower bound of the integral using $0$. If we instead use $\alpha_1$ for the lower bound we will get a well defined result for $d=3$ given by:
\begin{equation}
\begin{split}
    &\frac{dM}{dt}\bigg\vert_{d=3} \approx \frac{N_b}{2\pi\beta^2} \frac{\pi^4}{15}\left( \frac{r_s^2f''_{ext}(r_s)}{2} \right)\int_{\alpha_1}^\infty \left(\frac{\beta\delta r f''_{ext}(r_s)}{2}\right)^{-2-2\alpha_{\ell}^{2}}\alpha_{\ell}^{-1}d\alpha_{\ell}\\
    &=\frac{N_b}{2\pi\beta^2} \frac{\pi^4}{15}\left( \frac{r_s^2f''_{ext}(r_s)}{2} \right)\left[-\frac{Ei\left[-2\alpha_1^2\ln\left( \frac{\beta\delta r f''_{ext}(r_s)}{2} \right)\right]}{2\left(\frac{\beta\delta r f''_{ext}(r_s)}{2}\right)^2}\right], \\
    \end{split}
\end{equation}
where $Ei(x)$ is the exponential integral function. The leading order contribution for $\alpha_{1}\ll 1$ expansion gives the following evaporation rate:
\begin{equation}
    \frac{dM}{dt}\bigg\vert_{d=3}\approx \frac{N_b}{2\pi\beta^2} \frac{\pi^4}{15}\left( \frac{r_s^2f''_{ext}(r_s)}{2} \right)\left[\frac{-\gamma-\ln\left[2\alpha_1^2\ln\left(\frac{\beta\delta r f''_{ext}(r_s)}{2}\right)\right]}{2\left(\frac{\beta\delta r f''_{ext}(r_s)}{2}\right)^2}\right],
\end{equation}
where $\gamma\approx 0.577$ is the Euler–Mascheroni constant. Note that for this expansion to make sense $\alpha_1^2\ln\left(\beta\delta r f''_{ext}(r_s)\right)\ll 1$, this will be true if $r_s/L$ is sufficiently large (i.e. for very large AdS black holes):
\begin{equation}
\label{evapRateNearExtLargeAdSBHd3}
    \frac{dM}{dt}\bigg\vert_{d=3}\approx -\frac{N_b}{2\pi\beta^2} \frac{\pi^4}{15} \frac{6r_s^2}{L^2} \left[ \frac{\ln\left[\frac{2L^2}{3r_s^2}\ln\left(\frac{6\beta\delta r}{L^2}\right)\right]+\gamma}{2\left(\frac{6\beta\delta r}{L^2}\right)^2} \right].
\end{equation}

We compare the estimated evaporation rate to a numerical calculation of the full evaporation rate in Table \ref{largeBHtableapproxVsNum}. We find that the approximated result differs from the numerical result by a order one pre-factor, one can also check that the approximations will get better as $r_s/L$ becomes larger (we can see this graphically by comparing Figures (\ref{IntegrandPlotsbdel100r100} - \ref{IntegrandPlotsbdel100r10to50})).
\begin{table}[h!]
\begin{center}
\begin{tabular}{| c | c | c | c | c | c |} 
\hline
$d$ & 3 & 4& 6 & 8 & 10 \\ 
\hline
Numerical  & $3.48 \times 10^{-8}$  & $2.87\times 10^{-7}$ & $1.03\times 10^{-3}$ & $4.02\times 10^{1}$ & $6.25\times 10^{6}$  \\ 
\hline
Approximation & $3.40 \times 10^{-8}$  & $2.91\times 10^{-7}$ & $7.80\times 10^{-4}$ & $2.38\times 10^{1}$ & $2.79\times 10^{6}$  \\ 
\hline
$\mathcal{C}_{ext}=\frac{Numerical}{Approximation}$ & 1.02 & 0.99 & 1.32 & 1.69 & 2.24  \\ 
\hline
\end{tabular}
\end{center}
\caption{We fix $\frac{\beta\delta r}{r_s^2}=100$ and $r_s/L=100$. For different $d$ we numerically calculate the series defined in Eq. (\ref{SeriesRepofEvapRateNearExt}) and compare to the approximated evaporation rate we found in Eq. (\ref{evapRateNearExtLargeAdSBHdgeq4}) and Eq. (\ref{evapRateNearExtLargeAdSBHd3}) for $d\geq 4$ and $d=3$ respectively. We can see that in higher dimensions the approximation is not as good as it is in lower dimension but the results differ by an order one factor given by $\mathcal{C}_{ext}$. Furthermore, if one does similar calculations for larger values of $r_s/L$ we will find better agreement between the numerical result and approximated result.\label{largeBHtableapproxVsNum}}
\end{table}

For very small AdS black holes (or asymptotically flat black holes) in the near extremal regime we will not need to sum all the modes to infinity. We can get a rough estimate by simply computing the first term in the limit where $r_s/L\to 0$ we have\footnote{One can check that the ratio between the first and and second term in the series in the near extremal regime for very small AdS black holes will go as, $S_{tun,1}/S_{tun,2}\sim \left(\frac{\beta \delta r}{r_s^2}\right)^{\frac{2}{d-2}}\gg 1$. So the closer we are to the extremal regime smaller the sub-leading terms are compared to the first term. Furthermore, in higher dimensions we need to be closer to the extremal regime to similar errors as we might have in lower dimensions. We verify these statements with the results given in Table \ref{verysmallAdSExttable}.}:
\begin{equation}
\label{smallAdSbhNearExtEvapRate}
\begin{split}
    &\frac{dM}{dt}\approx \frac{N_b}{2\pi\beta^2} \left[\frac{\Gamma\left( 3+\sqrt{1+4\alpha_{1}^2} \right)Li_{3+\sqrt{1+4\alpha_1^2}}\left(1\right)}{x_{min,1}^{1+\sqrt{1+4\alpha_1^2}}} \right]  \\
    &= \frac{N_b}{2\pi\beta^2}\left[\frac{\Gamma\left(\frac{2(2d-3)}{d-2}\right)Li_{\frac{2(2d-3)}{(d-2)}}(1)}{\left((d-1)(d-2)^2\left(\frac{\beta \delta r}{r_s^2}\right)^2\right)^{\frac{d-1}{d-2}}}\right]=\frac{N_b}{2\pi\beta^2}\left[ \frac{\Gamma\left(\frac{2(2d-3)}{d-2}\right)Li_{\frac{2(2d-3)}{(d-2)}}(1)}{\left((d-1)(d-2)^2\right)^{\frac{d-1}{d-2}}} \right]\left(\frac{r_s^2}{\beta \delta r}\right)^{\frac{2(d-1)}{d-2}}. \\
    \end{split}
\end{equation}
In Table \ref{verysmallAdSExttable} we numerically verify that our estimation is valid when sufficiently close to the extremal regime.

\begin{table}[h!]
\begin{center}
\begin{tabular}{| c | c | c | c | c | c |} 
\hline
$\beta \delta r /r_s^2$ & $10^2$ & $10^4$& $10^8$ & $10^{16}$ & $10^{32}$ \\ 
\hline
$\mathcal{C}_{d=3}$ & $1.00$  & $1.00$ & $1.00$ & $1.00$ & $1.00$  \\ 
\hline
$\mathcal{C}_{d=4}$ & $1.01$  & $1.00$ & $1.00$ & $1.00$ & $1.00$  \\ 
\hline
$\mathcal{C}_{d=5}$ & $1.07$  & $1.00$ & $1.00$ & $1.00$ & $1.00$  \\ 
\hline
$\mathcal{C}_{d=6}$ & $1.33$  & $1.03$ & $1.00$ & $1.00$ & $1.00$  \\ 
\hline
$\mathcal{C}_{d=7}$ & $2.32$  & $1.16$ & $1.00$ & $1.00$ & $1.00$  \\ 
\hline
$\mathcal{C}_{d=8}$ & $6.40$  & $1.70$ & $1.03$ & $1.00$ & $1.00$  \\ 
\hline
$\mathcal{C}_{d=9}$ & $26.6$  & $3.82$ & $1.15$ & $1.00$ & $1.00$  \\
\hline
$\mathcal{C}_{d=10}$ & $153$  & $13.0$ & $1.64$ & $1.01$ & $1.00$  \\ 
\hline
\end{tabular}
\end{center}
\caption{We are setting $r_s/L=0$ (asymptotically flat black holes or very small AdS black holes). We are computing $\mathcal{C}_d$ which is the ratio between the numerical calculation of Eq. (\ref{SeriesRepofEvapRateNearExt}) divided by the approximated result given by Eq. (\ref{smallAdSbhNearExtEvapRate}) for spacetime dimension $d+1$. We can see that for larger values of $\beta$ the estimate for the evaporation rate using only $\ell=1$ mode becomes more precise. This is because of the for larger $\beta$ the $\ell=1$ mode is dominant compared to all the $\ell>1$ modes. \label{verysmallAdSExttable}}
\end{table}

Now that we have derived estimates for the evaporation rate it is useful to keep in mind that all the calculations we did in this subsection made the assumption that $\beta \omega_{min,\ell}\gg 1$. this implies that $\frac{\delta r}{r_s} \gg \left(\beta^2 f_{ext}''(r_s)\right)^{-1/2}$. We define the parameter $\Lambda$ as follows:
\begin{equation}
    \frac{\delta r}{r_s}=\frac{\Lambda}{\beta \sqrt{f''_{ext}(r_s)}}=\begin{cases}
    \frac{\Lambda L}{\beta\sqrt{2d(d-1)}}, & \text{for very large AdS BH in planar limit ($r_s/L\to \infty$)} \\
    \frac{\Lambda r_s}{\beta\sqrt{2(d-2)^2}}, & \text{for very small AdS (or asymptotically flat) BH},  \\
  \end{cases}
\end{equation}
where we require, $1\ll \Lambda \ll \beta\sqrt{f''_{ext}(r_s)}$. In terms of $\Lambda$ we express the evaporation rates of very large AdS black holes for $d\geq 4$ as:
\begin{equation}
\label{LargeAdSBHdgeq4EvapRate}
\begin{split}
   & \frac{dM}{dt}\bigg\vert_{d\geq 4}\approx \mathcal{A}^{large}_{d\geq 4} \left[\ln\left(\sqrt{\frac{d(d-1)}{2}}\frac{r_s}{L}\Lambda\right)\right]^{\frac{3-d}{2}} \frac{N_b}{\beta^2\Lambda^2}\left(\frac{r_s}{L}\right)^{d-3}\\
   &\mathcal{A}^{large}_{d\geq 4}=\frac{\pi^3}{30}\left(\frac{d(d-1)}{2}\right)^{\frac{d-3}{2}}\Gamma\left(\frac{d-3}{2}\right). \\
    \end{split}
\end{equation}
For very large AdS black holes for $d=3$ we have:
\begin{equation}
\label{LargeAdSBHd3EvapRate}
    \begin{split}
        \frac{dM}{dt}\bigg\vert_{d=3} \approx \frac{\pi^3}{30}\frac{N_b}{\beta^2\Lambda^2}\left[\ln\left(\frac{3r_s^2}{2L^2}\right)-\ln\left(\ln\left( \frac{\sqrt{3}r_s\Lambda}{L} \right)\right)-\gamma \right]\sim \frac{\pi^3}{30}\frac{N_b}{\beta^2\Lambda^2}\ln\left(\frac{3r_s^2}{2L^2}\right).
    \end{split}
\end{equation}
For very small AdS black holes (or asymptotically flat ones) we have:
\begin{equation}
\label{SmallAdSBHEvapRatee}
    \begin{split}
       &\frac{dM}{dt}\approx\frac{ \mathcal{A}_d^{small} N_b}{\beta^2\Lambda^{\frac{2(d-1)}{d-2}}} \\
       &\mathcal{A}_d^{small}=\frac{\left(\sqrt{2}(d-2)\right)^{\frac{2(d-1)}{d-2}}}{2\pi}\left[ \frac{\Gamma\left(\frac{2(2d-3)}{d-2}\right)Li_{\frac{2(2d-3)}{(d-2)}}(1)}{\left((d-1)(d-2)^2\right)^{\frac{d-1}{d-2}}} \right].\\
    \end{split}
\end{equation}
We will come back to the physical relevance of $\Lambda$ when discussing the ambiguities of ``fixing'' the screen a certain distance from the horizon in Section \ref{InfoEmAsScr}.

\section{Hayden-Preskill Decoding Criterion from Entanglement Wedge Reconstruction}
\label{PenningtonScrmabSec}
\subsection{Review of Pennington's Calculation}
As we discussed in the introduction, it was shown in \cite{Hayden:2007cs} that after the Page time a small amount of information thrown into a black hole could be reconstructed from subsequent Hawking radiation after the scrambling time scale. The works \cite{Penington:2019npb,Almheiri:2019psf} are able to reproduce this result in a holographic setting. The setup is to have the usual black hole in AdS which is dual to some CFT on the boundary. This is then supplemented by some type of absorbing boundary condition at the boundary which allows the radiation emitted by the black hole to be absorbed and stored. The radiation in the reservoir purifies the the black hole CFT state. There are two entanglement wedges in this scenario, one corresponds to the entanglement wedge of the black hole and the other is the entanglement wedge of the reservoir where radiation is absorbed. As the black hole evaporates these entanglement wedges have time dependence and it can be shown that information that is initially sitting in the entanglement wedge of the black hole a scrambling time in the past (assuming we are considering a time after the Page time) will end up in the entanglement wedge of the reservoir. This is equivalent to saying that information thrown into a black hole, after a Page time has elapsed, will re-emerge in the subsequent radiation after a scrambling time as is claimed in the Hayden-Preskill decoding protocol \cite{Hayden:2007cs}.      

In this section, we review some of the details of how this scrambling time scale appears in Pennington's calculations \cite{Penington:2019npb}. It comes from trying to find the location of a classical ``maximin'' surface in the spacetime of a spherically symmetric evaporating black hole (which happens to be a good approximation for where the quantum extremal surface is after a Page time has elapsed\footnote{It should be noted that we are actually interested in the quantum extremal surface which is found by applying the maximin prescription to a functional given by $\frac{A(\chi)}{4G_N}+S_{bulk}(\chi)$ . The calculation we are reviewing here finds the classical maximin surface, which is found by applying the maximin prescription to the area term ignoring the $S_{bulk}$ term. In \cite{Penington:2019npb}, Pennington argues that the classical maximin surface will only deviate slightly (even when one accounts for greybody factors) when the $S_{bulk}$ term is included and the quantum extremal surface will stabilize close to the classical maximin surface which lies on the lightcone. Throughout this paper we are going to assume that these arguments are still valid for our construction.}). The determination of the location of the surface eventually comes down to the following calculation. The first step is to start with a static spherically symmetric black hole metric of the form:
\begin{equation}
    ds^2=-f(r)dt^2+\frac{dr^2}{f(r)}+r^2d\Omega_{d-1}^2.
\end{equation}
Then one defines ingoing Eddington-Finkelstein coordinates $v=t+r_*$ where $dr=f(r)dr_*$. With some simple manipulations one arrives at the following metric:
\begin{equation}
    ds^2=-f(r)dv^2+2dvdr+r^2d\Omega_{d-1}^2.
\end{equation}
Upon doing this one approximates the metric of an evaporating black hole by introducing time dependence into $f$ by allowing the Schwarzschild radius $r_s$ to become time dependent (i.e. $r_s=r_s(v)$). One then considers radial null geodesics on this evaporating black hole spacetime. The radial coordinate $r_{lc}$ describing the trajectory of these null geodesics satisfy:
\begin{equation}
    \frac{dr_{lc}}{dv}=\frac{f(r_{lc})}{2}\simeq \frac{2\pi}{\beta}(r_{lc}-r_s).
\end{equation}
The right most expression comes from expanding $f(r_{lc})$ to first order and $\beta=T_H^{-1}=4\pi/f'(r_s)$. Define a coordinate $r'_{lc}=r_{lc}-r_s$ then we will find:
\begin{equation}
    \frac{dr'_{lc}}{dv}=\frac{2\pi}{\beta}r'_{lc}-\frac{dr_s}{dv}.
\end{equation}
Under the assumption that $dr_s/dv<0$ and approximately constant the equation can be integrated to find:
\begin{equation}
\label{LightRaySol}
    r_{lc}=r_s+Ce^{2\pi v/\beta}+\frac{\beta}{2\pi}\frac{dr_s}{dv},
\end{equation}
where $C$ is an integration constant. It is clear that when $C=0$, then $r_{lc}$ is constant (up to corrections caused by $dr_s/dv$ not being constant.) this defines the horizon of the evaporating black hole which is given by:
\begin{equation}
    r_{hor}=r_s\left( 1+\frac{\beta}{2\pi r_s}\frac{dr_s}{dv} \right)<r_s.
\end{equation}
Lets compute $dr_{hor}/dv$:
\begin{equation}
    \frac{dr_{hor}}{dv}=\frac{dr_s}{dv}+\frac{d\beta/dv}{2\pi}\frac{dr_s}{dv}=\frac{dr_s}{dv}+\mathcal{O}((dr_s/dv)^2)\sim \frac{dr_s}{dv}.
\end{equation}
With this we can compute $dr_{lc}/dv$:
\begin{equation}
    \frac{dr_{lc}}{dv}\simeq \frac{2\pi C}{\beta}e^{2\pi v/\beta}\left( 1-\frac{v}{\beta}\frac{d\beta}{dv} \right)+\frac{dr_s}{dv}+\mathcal{O}((dr_s/dv)^2).
\end{equation}
Assuming that $\left| \frac{v}{\beta}\frac{d\beta}{dv} \right|\ll 1$ we can solve for when $dr_{lc}/dv=0$ this occurs when\footnote{The length scale of $C$ was chosen by analyzing how far the expansion $f(r)$ near the horizon is valid to first order. In particular, it is not hard to see that $C\sim\frac{1}{\beta f''(r_s)}$. For small AdS black holes $f''(r_s)\sim r_s^{-2}$ (as noted by Pennington) and for large AdS black holes $f''(r_s)\sim L^{-2}$ (not discussed by Pennington), where $r_s$ and $L$ are the horizon and AdS radius respectively.}:
\begin{equation}
\label{ScramblingTime}
    v=v_0=-\frac{\beta}{2\pi}\ln\left( \frac{2\pi C}{\beta \left|\frac{dr_s}{dv}\right|} \right).
\end{equation}
To determine $|dr_s/dv|$, Pennington makes the assumption that Hawking quanta emitted by the black hole is assumed to be extracted sufficiently close to the horizon so that one can use the 2D Stefan-Boltzman law:
\begin{equation}
    \frac{dM}{dv}=\frac{c_{evap}\pi}{12\beta^2},
\end{equation}
where $c_{evap}=N_b+N_f/2$ where $N_b$ and $N_f$ are the number of bosonic and fermionic modes respectively. Using the first law of black hole thermodynamics the rate of energy loss can be related to $dr_s/dv$ the final result is:
\begin{equation}
    \left|\frac{dr_s}{dv}\right|=
\frac{4\beta\ell_p^{d-1}}{\Omega_{d-1}(d-1)r_s^{d-2}}\left| \frac{dM}{dv} \right|= \frac{c_{evap}\pi \ell_p^{d-1}}{3\beta(d-1)r_s^{d-2}\Omega_{d-1}}.
\end{equation}
This results in:
\begin{equation}
\label{PenningtonOriginalResults}
    v_0\simeq -\frac{\beta}{2\pi}\ln\left( \frac{Cr_s^{d-2}\Omega_{d-1}}{c_{evap}G_N} \right)\sim  \begin{cases}
    -\frac{\beta}{2\pi}\ln\left(\frac{r_s^{d-1}}{c_{evap}\ell_p^{d-1}}  \right), & \text{for non-extremal BH} \\
    -\frac{\beta}{2\pi}\ln\left( \frac{r_s}{c_{evap}\beta}\frac{r_s^{d-1}}{\ell_p^{d-1}} \right), & \text{for near extremal BH}.  \\
  \end{cases}
\end{equation}
So after the Page time, information thrown into the black hole reemerges after waiting for the time scale $|v_0|=t_{emerge}$ in Eq. (\ref{temerge}). Note that in the near extremal case the expression written down above is valid for small near extremal AdS black holes. For large near extremal AdS black holes $C\sim L^2/\beta$ so there will be some awkward $L$ dependence inside the Log. As we will see in the following sections, by properly understanding $c_{evap}$ for large AdS black holes, the length scale in the Log will come out to be $L$ instead of $r_s$.

\subsection{Information Emergence Time for AdS Schwarzschild Black Hole}
Using our newly derived evaporation rate in Eq. (\ref{SchAdSBHevaprate}) along with the first law of black hole thermodynamics and the area law for entropy of a black hole we will get:
\begin{equation}
\begin{split}
   & \left| \frac{dr_s}{dt} \right|=\frac{4\beta\ell_p^{d-1}}{(d-1)\Omega_{d-1}r_s^{d-2}}\left|\frac{dM}{dt}\right|=\frac{4\mathcal{A}_d}{(d-1)\Omega_{d-1}}\frac{N_b \ell_p^{d-1}}{\beta r_s^{d-2}}\left(\frac{r_s^2}{\delta r \beta}\right)^{\frac{d-1}{2}}\\
   &\sim \frac{N_b \ell_p^{d-1}}{\beta r_s^{d-2}}\left(\frac{r_s^2}{\delta r \beta}\right)^{\frac{d-1}{2}}.\\
    \end{split}
\end{equation}
To avoid clutter in our expressions we drop $\Omega_{d-1}$ and other dimensionless factors. Plugging this into Eq. (\ref{temerge}) we find for non-extremal black holes:
\begin{equation}
\label{ModifedPennScramNonExt}
    t_{emerge}\sim \frac{\beta}{2\pi}\ln\left( \frac{(\delta r \beta )^{\frac{d-1}{2}}}{N_b \ell_p^{d-1}} \right).
\end{equation}
For very large AdS Schwarzschild black holes ($r_s\gg L$) and the inverse temperature goes as $\beta\sim L^2/r_s$. Plugging this into Eq. (\ref{ModifedPennScramNonExt}) we find that Pennington's scrambling time scale results in:
\begin{equation}
\label{LargeNonExtScr}
    t_{emerge} \sim \frac{\beta}{2\pi}\left[\ln\left( \frac{L^{d-1}}{N_b\ell_p^{d-1}} \right)-\frac{d-1}{2}\ln\left( \frac{ r_s}{\delta r} \right)\right],
\end{equation}
where we assume that $L/\ell_p\gg r_s/\delta r$. The interesting thing to note here is that the leading order term is not the the Log of the entropy of the horizon of the black hole. It is actually the entropy of a small cell on the horizon which has the size of the AdS radius $L$. We can do a similar calculation for very small AdS black holes ($r_s\ll L$) in this case $\beta\sim r_s$ and we will obtain a more familiar result that Pennington got up to a Log correction that depends on where we place our absorptive screen:
\begin{equation}
     t_{emerge} \sim \frac{\beta}{2\pi}\left[\ln\left( \frac{r_s^{d-1}}{N_b\ell_p^{d-1}} \right)-\frac{d-1}{2}\ln\left( \frac{ r_s}{\delta r} \right)\right].
\end{equation}
As we can see from Eq. (\ref{LargeNonExtScr}), by understanding the explicit dependence of $c_{evap}$ on $\beta$ we find that $t_{emerge}$ contains the Bekenstein-Hawking entropy of a cell on the horizon of characteristic length $L$ inside the Log. This reasonable and consistent with the scrambling time discussed in \cite{Sekino:2008he} for large AdS black holes dual to large $N$ gauge theories\footnote{Recall that the ratio $(L/\ell_p)^{d-1}\sim N^2$ \cite{Rangamani:2016dms}.}.

\subsection{Information Emergence Time for Near Extremal AdS RN Black Hole}
Now lets consider what happens for near extremal AdS RN black holes. We can compute $|dr_s/dt|$ using the first law up to some dimensionless pre-factors we have:  
\begin{equation}
    \left|\frac{dr_s}{dt}\right|\sim \frac{\beta \ell_p^{d-1}}{r_s^{d-2}}\left|\frac{dM}{dt}\right|.
\end{equation}
We can compute $|dr_s/dt|$ using the evaporation rates in Eqs. (\ref{LargeAdSBHdgeq4EvapRate})-(\ref{SmallAdSBHEvapRatee}). We then plug these into Eq. (\ref{temerge}) and obtain the following results.\\

\textbf{Case 1, small AdS black hole: $r_s\ll L$}

In this case we have:
\begin{equation}
    t_{emerge}\sim \frac{\beta}{2\pi}\ln\left(\frac{r_s^2}{\beta^2\left|\frac{dr_s}{dt}\right|}\right) \sim  \frac{\beta}{2\pi}\left[\ln\left( \frac{r_s}{\beta}\frac{r_s^{d-1}}{N_b\ell_p^{d-1}} \right)+\frac{2(d-1)}{(d-2)}\ln\left(\Lambda\right)\right],
\end{equation}
where $1 \ll\Lambda \ll \beta/r_s$. \\

\textbf{Case 2, large AdS black hole: $r_s\gg L$}

In this case we have:
\begin{equation}
    \frac{2\pi t_{emerge}}{\beta}\sim \ln\left( \frac{L^2}{\beta^2\left|\frac{dr_s}{dt}\right|} \right) \sim \begin{cases}
    \ln\left(\frac{r_s}{\beta}\frac{L^2}{N_b\ell_p^2} \right)+\ln\left(\frac{\Lambda^2}{\ln(3r_s^2/L^2)}\right), & d=3 \\
 \ln\left(\frac{r_s}{\beta}\frac{L^{d-1}}{N_b\ell_p^{d-1}} \right)+\ln\left( \Lambda^2\left[\ln\left(\sqrt{\frac{d(d-1)r_s^2\Lambda^2}{2L^2}}\right)\right]^{\frac{d-3}{2}} \right) , & d\geq 4,  \\
  \end{cases}  \\
\end{equation}
where $1\ll \Lambda \ll \beta/L$.

If we make the assumption that $\Lambda$ has no additional $\beta$ dependence then we see the for small AdS black hole the first term matches what Pennington had. For large AdS black holes we again see that $L$ instead of $r_s$ appears in the leading order Log term. In the next section we will discuss an ambiguity that $\Lambda$ presents us with for near extremal black holes which is related to where we place our absorptive screen.  

\subsection{Information Emergence Time as Scrambling Time}
\label{InfoEmAsScr}
In the previous sections we found that the time scale after which information re-emerges for AdS Schwarzschild black holes is given as:
\begin{equation}
\label{nonextReemergenceTime}
    t_{emerge}\sim \begin{cases}
    \frac{\beta}{2\pi}\left[\ln\left( \frac{r_s^{d-1}}{N_b\ell_p^{d-1}} \right) -\frac{d-1}{2} \ln\left( \frac{r_s}{\delta r} \right) \right], & r_s\ll L \\
    \frac{\beta}{2\pi}\left[\ln\left( \frac{L^{d-1}}{N_b\ell_p^{d-1}} \right) -\frac{d-1}{2} \ln\left( \frac{r_s}{\delta r} \right) \right], & r_s\gg L.  \\
  \end{cases}
\end{equation}
For near extremal AdS RN black holes ($d\geq 4$) we have:
\begin{equation}
    t_{emerge}\sim \begin{cases}
    \frac{\beta}{2\pi}\left[\ln\left( \frac{r_s}{\beta}\frac{r_s^{d-1}}{N_b\ell_p^{d-1}} \right) + \frac{d-1}{d-2}\ln\left( \Lambda^2 \right)+... \right], & r_s\ll L \\
    \frac{\beta}{2\pi}\left[\ln\left( \frac{r_s}{\beta}\frac{L^{d-1}}{N_b\ell_p^{d-1}} \right) + \ln\left(\Lambda^2 \right)+... \right], & r_s\gg L,  \\
  \end{cases}
\end{equation}
where the ``...'' stand for double Log terms which we did not explicitly write. In the case of AdS Schwarzschild black holes we should assume the following hierarchy of scales that $\ell_p \ll \delta r\ll r_s$. By doing this it is clear that the dependence on $\delta r$ for the re-emergence time is sub-leading to the first term in the limit where $\ell_p\to 0$. We can reasonably identify $t_{emerge}$ with the scrambling time scales discussed in \cite{Hayden:2007cs,Sekino:2008he}. 

The case of a near extremal AdS RN black holes is more subtle. For near extremal AdS RN black holes we have an additional length scale that we did not have for the AdS Schwarzschild case. This length scale is $\beta$ and it causes problems when we try to decide on where the screen should be placed. To understand the issue, recall that we introduced $\Lambda$ through the following definition which relates it to $\delta r$:
\begin{equation}
    \Lambda=\frac{\delta r}{r_s} \beta \sqrt{f''_{ext}(r_s)},
\end{equation}
where we required that $1\ll \Lambda \ll \beta \sqrt{f''_{ext}(r_s)}$. The issue is that there are a number of choices we can make for the $\beta$-dependence of $\Lambda$. In Pennington's paper it is suggested that we extract radiation at a fixed distance from the horizon. There are at least two natural ways to do this. 

The first way is to set the radial coordinate distance from the horizon, $\delta r$, to some constant that does not depend explicitly on $\beta$. Then it is clear that $\Lambda\sim \beta/r_s$. In this case we would have results that look like:
\begin{equation}
    t_{emerge}\sim \begin{cases}
    \frac{\beta}{2\pi}\left[\ln\left( \left(\frac{\beta}{r_s}\right)^{\frac{d}{d-2}}\frac{r_s^{d-1}}{N_b\ell_p^{d-1}} \right) + \frac{d-1}{d-2}\ln\left( \frac{\delta r^2}{r_s^2} \right) \right], & r_s\ll L \\
    \frac{\beta}{2\pi}\left[\ln\left( \frac{\beta}{r_s}\frac{L^{d-1}}{N_b\ell_p^{d-1}} \right) + \ln\left(\frac{\delta r^2}{L^2} \right) \right], & r_s\gg L.  \\
  \end{cases}
\end{equation}
These results are at odds with what Pennington has for near extremal AdS black holes and also with the literature \cite{Leichenauer:2014nxa,Brown:2018kvn} which discusses the scrambling time for near extremal black holes. In particular, the main difference is how $\beta$ appears in the Log. One should expect $\beta$ to appear in the denominator rather than the numerator. This suggests that fixing an absorptive screen at a constant coordinate distance will yield a re-emergence time that is much longer than the scrambling time, $\frac{\beta}{2\pi} \ln (S-S_{ext})$.

Now consider the second way, which is to fix the proper radial distance from the screen to the horizon. Then we can show $\delta r \sim l_{prop}^2/\beta$, where $l_{prop}$ is the proper radial distance between the screen and horizon \footnote{To see this consider the proper radial length from the horizon to a point $\delta r$ from the horizon this is given by the an integral $l_{prop}=\int_{r_s}^{r_s+\delta r}\frac{dr}{\sqrt{f(r)}}$, for $\delta r \ll \min\{r_s,L\}$ we can expand to first order and do the integral to find $\delta r\sim \frac{l_{prop}^2}{\beta}$ }. By doing this, we see that $\Lambda$ will have no additional dependence on $\beta$ and we can write:

    \begin{equation}
\label{temergefixdelta}
    t_{emerge}\sim \begin{cases}
    \frac{\beta}{2\pi}\left[\ln\left( \frac{r_s}{\beta}\frac{r_s^{d-1}}{N_b\ell_p^{d-1}} \right) + \frac{2(d-1)}{d-2}\ln\left( \frac{l_{prop}^2}{r_s^2} \right) \right], & r_s\ll L \\
    \frac{\beta}{2\pi}\left[\ln\left( \frac{r_s}{\beta}\frac{L^{d-1}}{N_b\ell_p^{d-1}} \right) + \ln\left(\frac{l_{prop}^2}{L^2} \right) \right], & r_s\gg L,  \\
  \end{cases}
\end{equation}
then we can be reasonably identify the information re-emergence time with the scrambling time for near extremal black holes. So the question is what we should be fixing, the coordinate distance or proper distance, or perhaps something else? We believe the answer lies in the idea of fixing the energy scale of our effective theory on the screen. We know in the AdS/CFT correspondence the radial direction in the bulk corresponds to the energy scale of the CFT on the boundary. So by fixing the energy scale we should unambiguously fix how $\delta r$ scales with $\beta$. However, it is not clear exactly how the energy scale of the boundary theory depends on the radial distance. If it depends on the proper radial distance then we should fix the proper length between the horizon and screen. In discussions of the holographic renormalization group one usually considers metrics written in the form \cite{deBoer:2000cz}:
\begin{equation}
    ds^2=dz^2+e^{2z/L}\gamma_{ij}(z,x^i)dx^idx^j\label{metric},
\end{equation}
where $z$ is the radial direction in the bulk and $x^i$ are coordinates on the boundary and $\gamma_{ij}$ is the induced metric on a constant $z$ slice. The fixing of energy scales can be interpreted as the fixing of $z$. The way the metric is written suggests that $z$ is the proper radial length in the bulk. Therefore, it seems that fixing the proper length between the screen and horizon seems like a reasonable way to fix the energy scale, although this may not be valid for metrics that significantly differ from (\ref{metric}).

To summarize our discussion, we found that there are many ways to fix the $\beta$ dependence of $\Lambda$ (which is related to where the screen is placed). Depending on how $\delta r$ depends on $\beta$ we can get $t_{emerge}$ that may or may not resemble the scrambling time for near extremal AdS black holes. In particular, we find that by fixing the proper radial distance between the horizon and absorptive screen we get an information emergence time that is consistent with the scrambling time for near extremal AdS black holes. We suggested that fixing the proper radial distance between the screen and horizon can be interpreted as fixing the energy scale of the theory on the screen. We also find an additional sub-leading Log term which contains information on exactly where the screen absorbs radiation (which should not explicitly depend on $\beta$). It is interesting to note that for large AdS black holes it is not the entropy of the entire horizon that goes into the Log but instead the entropy of a cell on the horizon of characteristic length $L$. This is reasonable if we recall that large AdS black holes are dual to large $N$ gauge theories with $N^2\sim L^{d-1}/\ell_p^{d-1}$\cite{Rangamani:2016dms}.

\section{Discussion of the Physics of the Screen}

\subsection{Absorptive Screen as a Thin Shell of Matter}
\label{RigourousApproachtoGBF}
As we have demonstrated, the effect of extracting Hawking radiation near the horizon of a black hole generally has non-trivial consequences for the evaporation rate. In this work we adopted a model which extracted radiation close to the horizon using a perfectly absorbing screen that would absorb any Hawking radiation that gets to it. The rate at which energy was being absorbed by the screen for each angular momentum mode is captured through the generalized greybody factor. We did not rigorously compute this factor but instead proposed models that would capture the essential behaviour of the generalized greybody factor near the horizon. Here we will discuss a way to calculate the generalized greybody factor by treating the screen as an interface which patches an interior and exterior solution to the Einstein equations. 

\begin{figure}[h!]
\centering
\includegraphics[width=150mm]{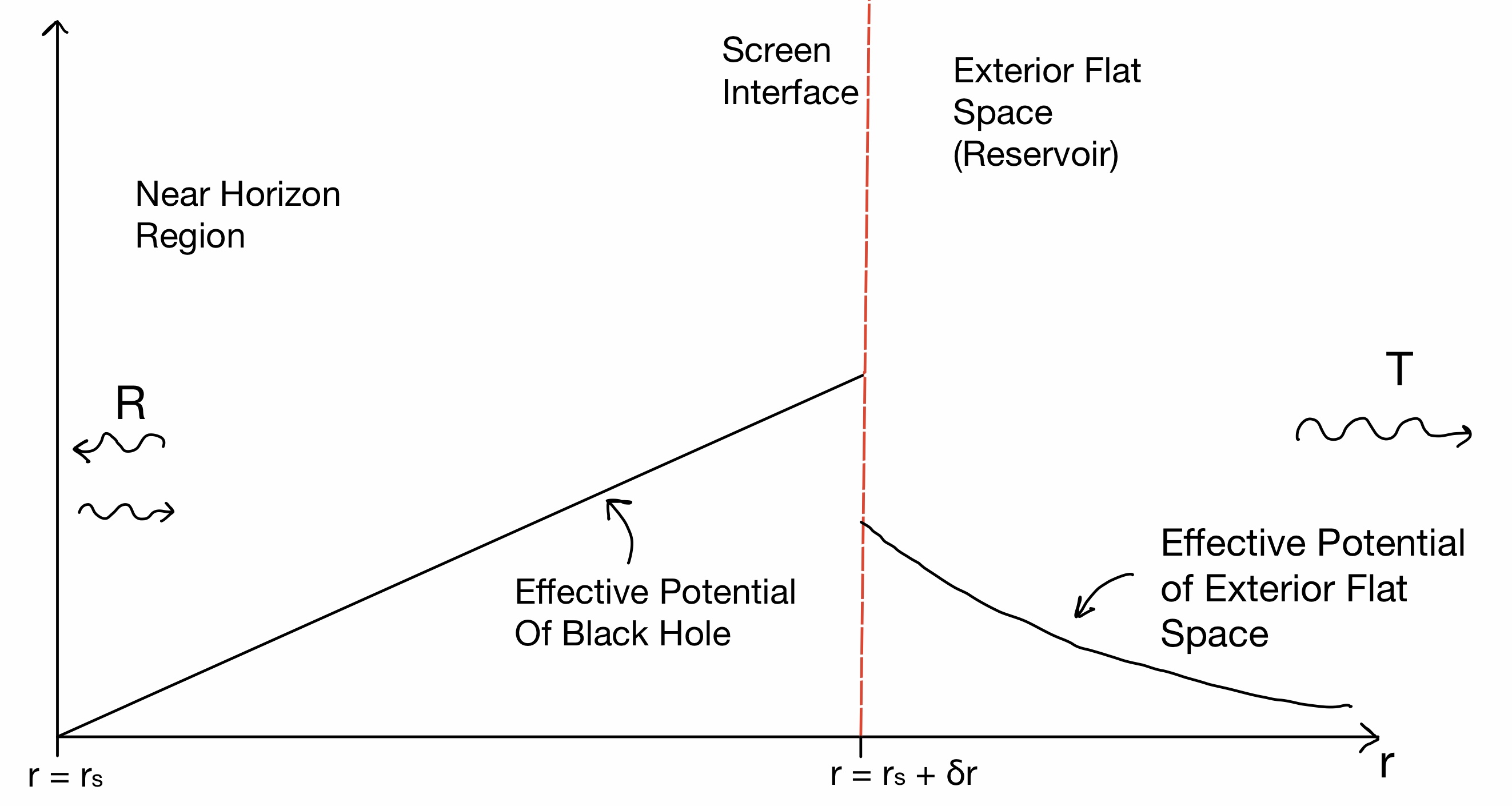}
\caption{Above is a depiction of the potential that we are considering to emulate an absorptive screen placed at $r=r_s+\delta r$ depicted by the dotted red line. We keep the effective potential the same as the black hole up until we get to the screen interface. We then transition to a potential for a flat space which will act as a reservoir for the extracted Hawking radiation. Close to the horizon the solution takes on the form of in-going and out-going plane waves. We normalize the outgoing wave near the horizon to unity and the amplitude of the in-going plane wave is $R$. The absorptive screen boundary condition is enforced by only allowing outgoing plane waves in the flat region with amplitude $T$. We patch the solutions and uniquely determine $T$ and $R$ by requiring continuity of the solution and its derivative at the screen interface. Then the generalized greybody factor is defined by $|T|^2$.          \label{ScreenPotentialFig} }
\end{figure}

In this picture, the screen is not really literally absorbing radiation, it is acting as an interface between the interior spacetime containing the black hole and an exterior ``reservoir'' spacetime which collects and stores the radiation emitted by the interior black hole. Assuming that the interior and exterior spacetimes are spherically symmetric, the scalar perturbations that propagate in this spacetime would satisfy the following radial wave equation:
\begin{equation}
    \frac{d^2\psi}{dr_*^2}+\left(\omega^2-V_{screen,\ell}\right)\psi=0,
\end{equation}
where $V_{screen,\ell}$ is the effective potential defined in a piece-wise manner in terms of the interior and exterior spacetimes:
\begin{equation}
\label{ScreenEffectivePotential}
    V_{screen,\ell}=\begin{cases}
   V_{int,\ell}(r), & r_s\leq r\leq r_s+\delta r \\
   V_{ext,\ell}(r), & r>r_s+\delta r.  \\
  \end{cases}
\end{equation}
The basic idea behind this is that we want to keep the spacetime unchanged until we arrive at the absorptive screen. The process of the screen ``absorbing'' radiation at $r+\delta r$ can be thought of a gluing an asymptotically flat region, just behind the screen and letting the wave ``escape to infinity''\footnote{The region behind the interface that we are gluing does not necessarily have to be an asymptotically flat space it could be more general. We choose an asymptotically flat space since the wave escaping to infinity would be the analogue of a purely absorptive boundary condition for the screen. One is also free to glue another asymptotically AdS space behind the screen. We will discuss this perspective in Section \ref{NEC c-thm discussion}.} as depicted in Figure \ref{ScreenPotentialFig}. 
To find the fraction of radiation ``absorbed'' by the screen (i.e. the generalized greybody factor) we would solve the wave equation in each region. In the interior region where $r\in (r_s,r_s+\delta r)$ the general solution will be some linear combination of two independent solutions:
\begin{equation}
    \psi_{I,\ell}(r_*)=c_{1,\ell}f_\ell(r_*)+c_{2,\ell}g_\ell(r_*).
\end{equation}
By analyzing the solution near the horizon we will find that they take on the form of plane waves and normalize the outgoing wave to unity (i.e. we start with outgoing Hawking radiation) this will fix some type of relation between $c_{1,\ell}$ and $c_{2,\ell}$. In the exterior region where the potential goes to zero far from the screen the solution should be purely outgoing plane wave (i.e.  absorptive screen boundary condition):
\begin{equation}
    \psi_{II,\ell}(r_*)=T_{\ell} e^{i\omega r_*}.
\end{equation}
We have 2 unknowns left now, namely $T$ and one of the coefficients of the solution of the interior region which will represent how much of the wave is reflected back. We can fix these by requiring the solution and its first derivative at $r=r_s+\delta r$ be continuous. This will fix $T_{\ell}$ uniquely. The generalized greybody factor is then defined by the the amplitude square of the transmission coefficient:
\begin{equation}
    \gamma_{\ell}(\omega,\delta r)=|T_{\ell}(\omega,\delta r)|^2.
\end{equation}
The procedure we outlined above would be a more rigorous way to find the generalized greybody factor. As one can imagine doing this analytically for any choice of $\delta r$ would be difficult, however the procedure we just outlined can be implemented numerically to find the exact behaviour of the generalized greybody factors. We expect that the generalized greybody factors to mimic the behaviour of the idealized models we analyzed in this paper at least in the limit where $\delta r\ll \min\{r_s,L\}$. It would be interesting to see how this method of extracting radiation at a finite distance from the horizon compares to other models that have been proposed to extract radiation from AdS black holes. For example, one could move the screen further from the horizon and ask how the generalized greybody factor at infinity (which is really just a greybody factor now) compares to greybody factors of models that use the evaporon \cite{Rocha:2009xy,Rocha:2008fe} to absorb energy from the black hole. 

\subsection{Null Energy Condition for the Screen and the Holographic c-Theorem}
\label{NEC c-thm discussion}
Recall that in Section \ref{introsec} of this work we wanted to view the absorptive screen near the horizon as a coarse-grained version of the conformal boundary (with absorbing boundary conditions). The goal of this discussion is to elaborate on this idea in the context of the holographic $c$-theorem and the role that the null energy condition plays in its formulation. 

The idea of the radial direction in the bulk being a measure of the energy scale of the dual boundary theory is formalized by discussing holographic $c$-theorems \cite{Freedman:1999gp,Myers:2010xs,Myers:2010tj}. When discussing $c$-theorems one usually considers two $d$-dimensional CFTs, one has a central charge $c_{UV}$ and the other has a central charge $c_{IR}$ where $c_{IR}<c_{UV}$. These two CFTs are assumed to be connected by an RG flow which starts from a UV fixed point and flows towards an IR fixed point. One can define a monotonic $c$-function which measures the effective degrees of freedom of the of the coarse-grained theory along the RG flow between the fixed points. If the two CFTs are holographic, one can make use of the AdS/CFT duality to construct a holographic $c$-function in terms of quantities defined in a $d+1$-dimensional gravity theory with matter. A central aspect of the construction relies on matter in the bulk satisfying the null energy condition (NEC). In particular, if one chooses appropriate coordinates so that the ``radial'' direction identifies the energy scales along the RG flow, then one needs the radial NEC to be satisfied in order to construct a monotonic $c$-function. Due to this fact, we will mainly focus on analyzing the radial NEC for the matter on the screen. Doing this we will provide a heuristic picture of how the effective degrees of freedom on the screen change as the screen is moved radially in the bulk.

To make things concrete, we will assume the interior spacetime, enclosed by the screen interface, is that of a $d+1$-dimensional AdS Schwarzschild black hole with a line element of the form:
\begin{equation}
\begin{split}
    &ds_-=g_{\mu\nu}^{-}dx^\mu dx^\nu=-f_-(r)dt^2+\frac{dr^2}{f_-(r)}+r^2d\Omega_{d-1}^2,\\
    &f_-(r)=1+\frac{r^2}{L_-^2}-\left(\frac{r_H}{r}\right)^{d-2}\left( 1+\frac{r_H^2}{L_-^2} \right),\\
    \end{split}
\end{equation}
where $r_H$ is the radial coordinate of the horizon, $L_-$ is the AdS radius for the solution. The ``$-$'' subscripts and superscripts mean we are dealing with quantities within the region enclosed by the screen. The exterior spacetime will be a pure AdS space with an AdS radius $L_+$ (``+'' superscripts and subscripts denote quantities in the exterior). The line element will be given by:
\begin{equation}
\begin{split}
    &ds_+=g_{\mu\nu}^{+}dx^\mu dx^\nu=-\Delta(r_0)f_+(r)dt^2+\frac{dr^2}{f_+(r)}+r^2d\Omega_{d-1}^2,\\
    &f_+(r)=1+\frac{r^2}{L_+^2},\\
    \end{split}
\end{equation}
where $r_0$ is the radial coordinate where the screen is placed. The lapse function $\Delta(r_0)=f_-(r_0)/f_+(r_0)$ ensures that the induced metric on either side of the screen is the same. Using the formalism described in Appendix \ref{DerivationOfShellEnergy} it can be shown that the stress energy tensor of the screen is given by Eq. (\ref{StressEnergy2}) and it resembles the stress energy tensor of a perfect fluid in $d$-dimensions with an energy density, $\rho$, and pressure, $p$ given by the following expressions:
\begin{equation}
    \begin{split}
    \label{PressureAndEnergyDensity}
         &\rho=\frac{(d-1)\left( f_-(r_0)^{1/2}-f_+(r_0)^{1/2} \right)}{8\pi r_0},\\
    &p=\frac{1}{16\pi r_0}\left[ 2(d-2)\left(f_+(r_0)^{1/2}-f_-(r_0)^{1/2}\right)+r_0\left(\frac{f'_+(r_0)}{f_+(r_0)^{1/2}}- \frac{f'_-(r_0)}{f_-(r_0)^{1/2}} \right) \right].
    \end{split}
\end{equation}
To summarize, we see that the patching of an interior black hole solution to an exterior AdS solution requires the screen to have a stress energy tensor of a $d$-dimensional perfect fluid with energy density and pressure given in Eq. (\ref{PressureAndEnergyDensity}). In Appendix \ref{ScreenNEC} we found the radial NEC translates to the screen having a positive energy density, $\rho \geq 0$. It turns out that the expression for the energy density of the screen can be positive only if $L_+\geq L_-$. Furthermore, the closest the screen can get to the horizon before the radial NEC is violated is given by:
\begin{equation}
\label{criticalrad2}
    r_c=r_H\left(\frac{1+\frac{L_-^2}{r_H^2}}{1-\frac{L_-^2}{L_+^2}}\right)^{1/d}.
\end{equation}
For any screen position $r_0>r_c$ the screen will have a positive energy density and the radial NEC is satisfied. Now consider holding $r_H$ and $L_-$ fixed and define $\mathcal{R}=L_+/L_-$. We will allow $\mathcal{R}$ to vary by changing the value of $L_+$. When $\mathcal{R}=1$ we know $r_c\to \infty$ so the screen has to be sitting at the conformal boundary in order to satisfy the radial NEC. If we increase $\mathcal{R}$ the screen is allowed to move deeper into the bulk. Now recall the standard dictionary in AdS/CFT which states that the AdS radius in Planck units is related to the effective number of degrees of freedom of the dual CFT \cite{Rangamani:2016dms}: 
\begin{equation}
    \left(\frac{L}{\ell_p}\right)^{d-1}\sim c_{eff}.
\end{equation}
Under the assumption that the screen is holographic we have a way to view the ratio $\mathcal{R}$ in terms of $c_{eff}$:
\begin{equation}
    \mathcal{R}^{d-1} \sim \frac{c^+_{eff}}{c^-_{eff}}\geq 1,
\end{equation}
where we defined $L_\pm^{d-1}/\ell_p^{d-1}\sim c_{eff}^{\pm}$. We view $\mathcal{R}$ as the ratio between the number of effective degrees of freedom of the screen and boundary theory. When the number of effective degrees of freedom of the screen equals the number effective degrees of freedom the boundary theory the screen must coincide with the boundary. If we coarse-grain the boundary theory (screen) the number of degrees of freedom on the screen are reduced and this corresponds to moving the screen deeper into the bulk. From this, we can heuristically see how satisfying the radial NEC for the screen gives rise to a monotonic decrease in the effective number of degrees of freedom on the screen as it is moved closer to the horizon of the black hole.  

\subsection{The Null Energy Condition and Black Hole Mining}
\label{MiningSection}
The idea of changing the evaporation rate of a black hole by extracting radiation near the horizon has also been discussed in the context of black hole mining \cite{1983GReGr..15..195U,Lawrence:1993sg,Frolov:2000kx}. In particular, Brown suggests that energy conditions (most notably the null energy condition) impose constraints on how quickly one can extract radiation from the horizon \cite{Brown:2012un}. In the previous subsection, we found that satisfying the radial NEC at a finite distance from the horizon places a constraint on how close the screen is allowed to be to the horizon. The closest radial coordinate is given by $r_c$ in Eq. (\ref{criticalrad2}). Then $\delta r_{min}=r_c-r_H$ is given by:
\begin{equation}
    \delta r_{min}=\left[\left(1+\frac{L_-^2}{r_H^2}\right)^{1/d}-1\right]r_H\simeq \begin{cases}
    \left(\frac{L_-}{r_H}\right)^{2/d}r_H, & r_H\ll L_- \\
   \frac{1}{d}\left(\frac{L_-}{r_H}\right)^2r_H, & r_H\gg L_-,  \\
  \end{cases}   \\
\end{equation}
where the expression above is taken in the limit that $L_+ \to \infty$ so the exterior spacetime is asymptotically flat. An interesting observation  is that  $\delta r_{min}$ monotonically increases as the black hole evaporates. Due to this, we can see that for small AdS black holes the screen cannot be placed very close to the horizon, so we are not really mining very small AdS black holes with the screen\footnote{The reader may be concerned with the evaporation rate we derived in Eq. (\ref{SchAdSBHevaprate}) for the small AdS black hole regime. The equation was derived assuming the screen is placed close to the horizon, but respecting the radial NEC does not allow this. This should not be a particularly big issue since, up to an order one pre-factor,  the evaporation rate should go as $dM/dt\sim \beta^{-2}$ \cite{Mistry:2017ubm}. Which is consistent with Pennington's results \cite{Penington:2019npb} as well as ours.}. However, for very large AdS back holes the screen can be placed very close to the horizon. In this case it is interesting to ask how long it takes for a very large AdS black hole to transition to a small AdS black hole via screen mining. We estimate this time by setting $\delta r=\delta r_{min}$ in Eq. (\ref{SchAdSBHevaprate}) to get the following evaporation rate for very large AdS black holes:
\begin{equation}
\frac{dM}{dt}\sim \frac{N_b}{L_-^2}\left(\frac{r_H}{L_-}\right)^{\frac{4d+3}{2}},   \end{equation}
We use the relation between the mass and horizon radius of very large AdS black holes, given by $M\sim r_H^d/(L_-^2\ell_p^{d-1})$, to replace the derivative of mass with derivative of the horizon radius. We integrate the equation to estimate the duration of time elapsed for the large AdS black hole with an initial radius of $r_H=r_s\gg L_-$ to evaporate to a black hole of radius $L_-$. We find:
\begin{equation}
\label{HPTransTime}
    \Delta t\sim \left(\frac{L_-}{\ell_p}\right)^{d-1}\left[1-\left(\frac{L_-}{r_s}\right)^{d+\frac{3}{2}}\right]\frac{L_-}{N_b}\approx \left(\frac{L_-}{\ell_p}\right)^{d-1}\frac{L_-}{N_b}.
\end{equation}
At leading order we find that the time it takes (in units of AdS radius) for a very large AdS black hole to transition to the small AdS black hole regime via screen mining is proportional the Bekenstein-Hawking entropy of an AdS cell. After the black hole enters the small regime the evaporation rate will mimic that of a black hole evaporating in asymptotically flat space (i.e. evaporation rate will go as $dM/dt\sim N_b \beta^{-2}$). It is difficult to directly compare our result for the evaporation rate of large AdS black holes with the results of Brown \cite{Brown:2012un} which are concerned with asymptotically flat black holes. However, we can see that once the black hole enters the small regime the bounds derived by Brown are not violated because the screen is far from the horizon (i.e. radial NEC only allows near horizon screen mining of very large AdS black holes). 

It is interesting to mention that the time scale in Eq. (\ref{HPTransTime}) we found using near horizon screen mining is agrees with the evaporation time scale found in an earlier work by Page \cite{Page:2015rxa}. Page's work considers large AdS black hole evaporation assuming absorptive boundary conditions at infinity. Having absorptive boundary conditions at infinity is analogous to placing our absorptive screen at infinity. The fact that the lifetimes in either case (i.e. near or far screen mining) are comparable to each other suggests that the lifetime of very large AdS black holes does not significantly change when mined by a screen obeying the radial NEC. 

So far, we have restricted ourselves to discussing the NEC for null vectors with only a radial component. This was primarily because of the connection between the radial NEC and discussions of the holographic $c$-theorem. One may ask what kind of constraints the the NEC gives if the null vectors are tangent to the screen (i.e. no radial component).  In Appendix \ref{ScreenNEC} we show that the screen violates the tangential NEC at any finite distance from the horizon. The violations of the tangential NEC become milder the further the screen is placed from the horizon and is actually saturated in the limit where the screen is sent to infinity. This is unsurprising as a screen composed of ordinary matter will not sit at  a fixed distance from the horizon, but rather would fall into the black hole. In order for it not to fall in the matter composing the screen must violate energy conditions. However, it is worth noting that the calculations we did, did not account for Hawking radiation being emitted from the black hole. It is well known that Hawking radiation violates energy conditions, which is why the area of the horizon decreases \cite{Ford:1995gb,Lesourd}. An interesting idea worth considering is whether the screen can be prevented from falling into the black hole by the pressure generated by the Hawking radiation emitted by the black hole. Naively, the pressure due to Hawking radiation will become larger the closer the screen gets to the horizon this may counteract the gravitational pull on the screen generated by the horizon.

\section{Conclusion and Future Prospects}
In this work, we investigated how the evaporation rate of AdS black holes change when radiation is absorbed near the horizon using an absorptive screen, which is motivated by the entanglement wedge reconstruction framework described by Pennington \cite{Penington:2019npb}. We used idealized toy models, motivated by physical arguments, which would capture the essential physics of radiation propagating towards the absorptive screen. We showed that by fixing the screen at a proper radial distance from the horizon, the re-emergence time for the information thrown into an AdS black hole is given by the expressions in Eq. (\ref{nonextReemergenceTime}) and Eq. (\ref{temergefixdelta}). For small AdS black holes (or asymptotically flat black holes) the expressions, at leading order, contain the Log of the entropy of the whole horizon. This is consistent with Pennington's calculations \cite{Penington:2019npb}. In contrast, however, for large AdS black holes, we find that the re-emergence time depends on the log of the entropy of an AdS cell on the horizon (rather than that of the entire horizon). Such a modification is reasonable and consistent with the scrambling time discussed in the work \cite{Sekino:2008he}.

In Sections \ref{RigourousApproachtoGBF} and \ref{NEC c-thm discussion}, we attempted to provide a more physical description of what governs wave propagation and internal physics of our putative screen. The interior spacetime enclosed by the screen contains the black hole, while an exterior asymptotically flat or AdS spacetime can represent the auxiliary system that could store radiation. The ``absorption'' of radiation by the screen would then be equivalent to radiation passing through the screen interface and escaping to infinity. The calculation of how radiation would be ``absorbed'' can be translated to a well-defined computation of greybody factors. We used Israel junction conditions to compute the stress associated with the screen. The requirement of the matter on the screen having a positive energy density (which comes from the radial NEC used to formulate a monotonic $c$-function in a holographic RG description)  sets a minimum distance for the screen from the black hole horizon if it is the dual description of a coarse-grained unitary boundary CFT. 

 The calculations done in this paper have been done from a gravitational perspective. In order to explore the ideas discussed in Sections \ref{RigourousApproachtoGBF} and \ref{NEC c-thm discussion} more rigorously it will be necessary to translate the gravitational picture we proposed to a coupled quantum system description. The $T\bar{T}$ formalism described in \cite{McGough:2016lol} will likely be an important ingredient and a good starting point for defining the screen theory. We would then couple the screen theory defined by the $T\bar{T}$ deformation to a holographic CFT describing an AdS bulk with a brick wall. We leave such a formulation to future work.

\acknowledgments

We would like to thank Geoffrey Pennington for helpful comments and discussions, as well as Adam Brown for bringing the issue of NEC and black hole mining to our attention (now discussed in Section \ref{MiningSection} with supplementary Appendices \ref{DerivationOfShellEnergy} and \ref{ScreenNEC}).   This work was supported by the University of Waterloo, Natural Sciences and Engineering Research Council of Canada (NSERC), and the Perimeter Institute for Theoretical Physics. Research at Perimeter Institute is supported in part by the Government of Canada through the Department of Innovation, Science and Economic Development Canada and by the Province of Ontario through the Ministry of Colleges and Universities.

\appendix

\section{Numerical Analysis of Evaporation Rate Series for AdS Schwarzschild Black Hole}
\label{EvapApproxAppendix}
In this appendix we will numerically compute the following series for an AdS Schwarzschild black hole:
\begin{equation}
\begin{split}
\label{Schexactser}
   & \sum_{\ell=1}^{\infty}\ell^{d-2}\left[ Li_2\left(e^{-x_{min,\ell}}\right)-x_{min,\ell}\ln\left(1-e^{-x_{min,\ell}}\right) \right]\\
   &x_{min,\ell}=\sqrt{4\pi\left(2\pi(d-1)+\frac{4\pi L^2 \ell(\ell+d-2)}{dr_s^2+(d-2)L^2}\right)\frac{\delta r}{r_s}}.\\
    \end{split}
\end{equation}
To evaluate the series numerically we need to fix $d$, $\delta r/r_s$, and $r_s/L$. Once we do this we will compare the result to our approximated expression given by:
\begin{equation}
\begin{split}
\label{approxser}
    &\sum_{\ell=1}^{\infty}\ell^{d-2}\left[ Li_2\left(e^{-x_{min,\ell}}\right)-x_{min,\ell}\ln\left(1-e^{-x_{min,\ell}}\right) \right]\\
    &\approx d(d-2)!\left(\frac{r_s^2}{4\pi\beta\delta r}\right)^{\frac{d-1}{2}}.\\
    \end{split}
\end{equation}
The approximated expression will differ from the numerical expression by a numerical pre-factor. In other words the numerical result can be written in the form:
\begin{equation}
\begin{split}
    &\sum_{\ell=1}^{\infty}\ell^{d-2}\left[ Li_2\left(e^{-x_{min,\ell}}\right)-x_{min,\ell}\ln\left(1-e^{-x_{min,\ell}}\right) \right]\\
    &= \mathcal{C}_d \left[d(d-2)!\left(\frac{r_s^2}{4\pi\beta\delta r}\right)^{\frac{d-1}{2}}\right],\\
    \end{split}
\end{equation}
where $\mathcal{C}_d$ is a numerical pre-factor which will change with $d$, $\delta r$, and $r_s/L$. We summarize our results in tables \ref{Tabled3delta10m4} - \ref{Tabled10delta10m4}. Each table fixes $d$ and $\delta r/r_s$ to some fixed value (specified in the caption for each table). Within the table we vary the size of the black hole $r_s/L$ (from $0$ corresponding to an asymptotically flat black hole to $1000$ corresponding to a very large AdS black hole) and numerically compute the series in Eq. (\ref{Schexactser}). We also compute the value for series as determined by our approximation given in Eq. (\ref{approxser}). We divide the numerical and approximate result to determine the pre-factor $\mathcal{C}_d$ that the two results differ by. 

\begin{table}[h!]
\begin{center}
\begin{tabular}{| c | c | c | c | c | c |} 
\hline
$r_s/L$ & 0 & 0.1& 1 & 10 & 1000 \\ 
\hline
Numerical  & $1.96\times 10^{2}$  & $2.02\times 10^{2}$ & $8.01\times 10^{2}$ & $6.15\times 10^{4}$ & $6.14\times 10^{8}$  \\ 
\hline
Approximation & $1.89\times 10^{2}$  & $1.95\times 10^{2}$ & $7.59\times 10^{2}$ & $5.72\times 10^{4}$ & $5.70\times 10^{8}$  \\ 
\hline
$\mathcal{C}_3=\frac{Numerical}{Approximation}$ & 1.04 & 1.04 & 1.06 & 1.08 & 1.08  \\ 
\hline
\end{tabular}
\end{center}
\caption{$d=3$ and $\delta r/r_s=10^{-4}$ \label{Tabled3delta10m4}}
\end{table}

\begin{table}[h!]
\begin{center}
\begin{tabular}{| c | c | c | c | c | c |} 
\hline
$r_s/L$ & 0 & 0.1& 1 & 10 & 1000 \\ 
\hline
Numerical  & $1.33\times 10^{8}$  & $1.38\times 10^{8}$ & $1.37\times 10^9$ & $4.11\times 10^{13}$ & $4.07\times 10^{23}$  \\ 
\hline
Approximation & $1.47\times 10^8$  & $1.53\times 10^8$ & $1.45\times 10^{9}$ & $4.12\times 10^{13}$ & $4.05\times 10^{23}$  \\ 
\hline
$\mathcal{C}_6=\frac{Numerical}{Approximation}$ & 0.91 & 0.90 & 0.95 & 1.00 & 1.01  \\ 
\hline
\end{tabular}
\end{center}
\caption{$d=6$ and $\delta r/r_s=10^{-4}$ \label{Tabled6delta10m4}}
\end{table}

\begin{table}[h!]
\begin{center}
\begin{tabular}{| c | c | c | c | c | c |} 
\hline
$r_s/L$ & 0 & 0.1& 1 & 10 & 1000 \\ 
\hline
Numerical  & $5.09\times 10^{17}$  & $5.38\times 10^{17}$ & $2.06\times 10^{19}$ & $1.66\times 10^{27}$ & $1.63\times 10^{45}$  \\ 
\hline
Approximation & $5.98\times 10^{17}$  & $6.32\times 10^{17}$ & $2.30\times 10^{19}$ & $1.69\times 10^{27}$ & $1.63\times 10^{45}$  \\ 
\hline
$\mathcal{C}_{10}=\frac{Numerical}{Approximation}$ & 0.85 & 0.85 & 0.90 & 0.98 & 1.00  \\ 
\hline
\end{tabular}
\end{center}
\caption{$d=10$ and $\delta r/r_s=10^{-4}$ \label{Tabled10delta10m4}}
\end{table}

\section{Power Law Behaviour of Generalized Greybody Factor for Near Extremal BH}
\label{PowerLawAppendix}
Here we present a way to get the power law behaviour for $\omega<\omega_{min,\ell}$ in Eq. (\ref{GeneralizedGBFNearExt}). We do this by analyzing the near horizon solution of the wave equation for an extremal black hole. We will begin by considering modes with $\ell\geq 1$.

The first thing we do is recall that the potential needs to be written in the tortoise coordinate $r_*$ which satisfies:

\begin{equation}
    r_*=\int\frac{dr}{f(r)}\simeq \int \frac{dr}{f_1(r-r_s)+\frac{f_2}{2}(r-r_s)^2}=\frac{1}{f'(r_s)}\ln\left[ \frac{(r-r_s)f''(r_s)}{2f'(r_s)+f''(r_s)(r-r_s)} \right]\leq 0,
\end{equation}
where $f_n=f^{(n)}(r_s)$. We can easily invert this and find:

\begin{equation}
    r-r_s=\frac{2f_1}{f_2(1-\exp(f_1r_*))}\to -\frac{2}{r_*f''_{ext}(r_s)},
\end{equation}
where in the last expression we take the extremal limit where $f_1\to 0$. Now that we have an expression for the near horizon tortoise coordinate we can analyze the wave equation which at leading order will read:

\begin{equation}
    \frac{d^2\psi}{dr_*^2}+\left[ \omega^2-\frac{2\ell(\ell+d-2)}{r_s^2f''_{ext}(r_s)}\frac{1}{r_*^2} \right]\psi=0.
\end{equation}
We can find the general solution to this equation can be written in terms of Bessel functions:

\begin{equation}
\begin{split}
   & \psi(r_*)=\sqrt{r_*}\left[A J_{\nu_{\ell}}\left( \omega r_* \right)+BY_{\nu_{\ell}}\left( \omega r_* \right)\right]\\
   &\nu_\ell=\frac{1}{2}\sqrt{1+\frac{8\ell(\ell+d-2)}{r_s^2f''_{ext}(r_s)}}.\\
    \end{split}
\end{equation}
We want a solution that goes to zero at $r_*=0$ this implies that $B=0$ and we get the following solution:

\begin{equation}
    \psi(r_*)=A\sqrt{r_*}J_{\nu}(\omega r_*).
\end{equation}
One can easily see that this solution for very small $r_*$ oscillates as a plane wave. However near the boundary it decays. It is the rate of decay that we are interested in. In particular, it is reasonable to assume that the shift from an oscillating function to a decaying function occurs near the classical turning point which is:

\begin{equation}
\begin{split}
    &r_*^{tp}(\omega)=-\left[\frac{2\ell(\ell+d-2)}{\omega^2r_s^2f''_{ext}(r_s)}\right]^{1/2}.\\
    \end{split}
\end{equation}
Consider the ratio:

\begin{equation}
T^2=\left|\frac{\psi(r_*)}{\psi(r_*^{tp})}\right|^2=\left|\frac{r_*}{r_*^{tp}} \right| \left|\frac{J_{\nu}(\omega r_*)}{J_{\nu}(\omega r_*^{tp})} \right|^2,
\end{equation}
where $r_{*}^{tp}\leq r_*\leq 0$. This gives a measure of how the amplitude of the solution decays in the non-classical region. We analyze the decay of the solution a distance $\delta r=r-r_s$ from the horizon in the classically forbidden region. To do this we need to consider $\omega\leq \omega_{min,\ell}$. We parameterize this in terms of $0 \leq \epsilon \leq 1$ and write $\omega=\epsilon \omega_{min,\ell}$. Then we can express $r_{*}^{tp}$ as:

\begin{equation}
    r_*^{tp}(\epsilon)=-\frac{2}{\epsilon \delta r f''_{ext}(r_s)}.
\end{equation}
We also set $r_*$ at the position of interest (i.e. where the absorbing surface is):

\begin{equation}
    r_*=-\frac{2}{\delta r f''_{ext}(r_s)}.
\end{equation}
Now we can express $T^2$ in terms of $\epsilon$:

\begin{equation}
\begin{split}
    &T^2(\epsilon) =\epsilon \left|\frac{J_{\nu_{\ell}}\left( \alpha_\ell \epsilon \right)}{J_{\nu_{\ell}}\left( \alpha_\ell \right)} \right|^2\\
    &\alpha_{\ell}= \frac{-2 \omega_{min,\ell}}{\delta r f''_{ext}(r_s)}=-\sqrt{\frac{2\ell(\ell+d-2)}{r_s^2f''_{ext}(r_s)}}  \\
    &\nu_\ell=\sqrt{\frac{1}{4}+\alpha_{\ell}^2}. \\
    \end{split}
\end{equation}
We can do a series expansion of $T^2$ in $\alpha_{\ell}$ to understand the power law behaviour we find:

\begin{equation}
   T^2=\epsilon \left|\frac{J_{\nu_{\ell}}\left( \alpha_\ell \epsilon \right)}{J_{\nu_{\ell}}\left( \alpha_\ell \right)} \right|^2\simeq \epsilon^{2\nu_{\ell}+1}\left[ 1+\frac{\alpha_\ell^2(1-\epsilon^2)}{2(1+\nu_\ell)}+\mathcal{O}\left(\alpha_{\ell}^4\right) \right]\sim \epsilon^{2\nu_{\ell}+1}.
\end{equation}
We will use this behaviour to model the generalized greybody factor for $\omega<\omega_{min,\ell}$. So now we have the following for near extremal black holes:

\begin{equation}
    \gamma_\ell(\omega,\delta r)=\Theta(\omega-\omega_{min,\ell})+\left(\frac{\omega}{\omega_{min,\ell}}\right)^{2\nu_{\ell}+1}\Theta\left( \omega_{min,\ell}-\omega \right).
\end{equation}
This gives the result in Eq. (\ref{GeneralizedGBFNearExt}).

Now we will consider the $\ell=0$ mode. In this case the leading order expansion of the effective potential near the horizon is:
\begin{equation}
    V_{\ell=0}(r)=V_0(r)=\frac{(d-1)\left[f_{ext}''(r_s)\right]^2}{4r_s}(r-r_s)^3+\mathcal{O}\left((r-r_s)^4\right).
\end{equation}
Note that to capture the leading order behavior of the effective potential for $\ell=0$ one must expand to third order. This is in contrast to the effective potential for $\ell\geq 1$ modes which. only required a second order expansion. As we will see this makes the $\ell=0$ modes distinct from the higher order modes. With some work we can show that the wave equation near the horizon takes the form:
\begin{equation}
    \frac{d^2\psi}{dr_*^2}+\left[\omega^2+\frac{2(d-1)}{r_sf''_{ext}(r_s)}\frac{1}{r_*^3}\right]\psi=0.
\end{equation}
Unlike the $\ell \geq 1$ case we cannot find the general solution of this equation in a closed form. We instead opt to solve the equation in two regimes (close to the horizon and close to the conformal boundary) and then patch the solutions at the turning point of the potential. 

Close to the horizon we have plane wave solutions. The outgoing plane wave normalized to one is given by:
\begin{equation}
    \psi_{I}(r_*)=e^{i\omega r_*}.
\end{equation}

In the classically forbidden region (where the amplitude of the solution will decay) we will have $\omega^2\ll \frac{2(d-1)}{r_sf''_{ext}(r_s)}\frac{1}{r_*^3}$. So the solution can be roughly found by solving:
\begin{equation}
     \frac{d^2\psi}{dr_*^2}+\frac{2(d-1)}{r_sf''_{ext}(r_s)}\frac{1}{r_*^3} \psi=0.
\end{equation}
The general solution will be given by Bessel functions of the first and second kind:
\begin{equation}
\begin{split}
   & \psi_{II}(r_*)=\frac{\sqrt{-r_*}}{\alpha}\left[ c_1J_1\left(\frac{2i\alpha}{\sqrt{-r_*}}\right)+c_2Y_1\left( \frac{2i\alpha}{\sqrt{-r_*}} \right) \right]\\
   &\alpha=\sqrt{\frac{2(d-1)}{r_sf''_{ext}(r_s)}}.\\
    \end{split}
\end{equation}
Just like for the $\ell\geq 1$ modes we impose the boundary condition that the solution vanish at the conformal boundary located at $r_*=0$. This gives the following solution the the forbidden region:
\begin{equation}
    \psi_{II}(r_*)=\frac{A\sqrt{-r_*}}{\alpha}\left[ J_1\left(\frac{2i\alpha}{\sqrt{-r_*}}\right)+iY_1\left( \frac{2i\alpha}{\sqrt{-r_*}} \right) \right].
\end{equation}
Patching the solutions in the two regions at the turning point $r_*^{tp}$ by requiring $\psi_I(r_*^{tp})=\psi_{II}(r_*^{tp})$ allows us to fix the constant $A$. To find the power law decay we analyze how the amplitude of the solution decays from the turning point to the screen in a similar manner to what we did for the $\ell\geq 1$ mode. In particular, the fraction of radiation that gets to the screen is given by:
\begin{equation}
\begin{split}
   & T_0^2=\left|\frac{\psi_{II}(r_*^{screen})}{\psi_{II}(r_*^{tp})}\right|^2\\
   &r_*^{screen}=\frac{-2}{\delta r f_{ext}''(r_s)} \\
   &r_*^{tp}(\omega)=-\left(\frac{2(d-1)}{\omega^2r_sf''_{ext}(r_s)}\right)^{1/3}. \\
    \end{split}
\end{equation}
We define $\omega=\epsilon \omega_{min,0}$ where $0\leq \epsilon \leq 1$ and:
\begin{equation}
\begin{split}
    &\omega_{min,0}=\eta \frac{\delta r f''_{ext}(r_s)}{2}\\
    &\eta = \left(\frac{(d-1)\delta r}{r_s}\right)^{1/2}\ll 1,\\
    \end{split}
\end{equation}
$\omega_{min,0}$ is the minimal frequency in which waves would reach the screen without encountering the angular momentum barrier for $\ell=0$. We can then write $T_0^2$ as:
\begin{equation}
    T_0^2=\epsilon^{2/3}\left| \frac{J_1\left(2i\eta \right)+iY_1\left(2i\eta \right)}{J_1\left( 2i\eta \epsilon^{1/3} \right)+iY\left( 2i\eta \epsilon^{1/3} \right)} \right|^2\simeq \epsilon^{4/3}\left[1+\mathcal{O}\left(\eta^2\right)\right]\sim \epsilon^{4/3}.
\end{equation}
So for the $\ell=0$ mode the generalized greybody factor in our toy model will be:
\begin{equation}
\label{GGBFleeq0}
    \gamma_0\left( \omega,\delta r \right)=\Theta\left( \omega-\omega_{min,0} \right)+\left(\frac{\omega}{\omega_{min,0}}\right)^{4/3}\Theta\left( \omega_{min,0}-\omega \right).
\end{equation}

\section{Integrals Describing Evaporation rate in Near Extremal Regime}
\label{IntegralsNearExt}
In this section we go over the assumptions to arrive at the series expression for the evaporation rate given by Eq. (\ref{SeriesRepofEvapRateNearExt}). We need to compute the integrals in Eq. (\ref{NearExtEvapRateInteg}).

We approximate the values of these integrals under the assumption that $\beta \omega_{min,\ell}\gg 1$ (i.e. we are sufficiently close to the extremal regime). Lets begin with the first term(s) in Eq. (\ref{NearExtEvapRateInteg}) which describes modes with $\omega\leq \omega_{min,\ell}$. The term(s) read: 
\begin{equation}
\begin{split}
    &\frac{N_b}{2\pi}\sum_{\ell=1}^\infty \omega_{min,\ell}^2N_\ell \int_0^1 \frac{\epsilon^{2(\nu_\ell+1)}}{e^{\beta\omega_{min,\ell}\epsilon}-1}d\epsilon=\frac{N_b}{2\pi}\sum_{\ell=1}^\infty \omega_{min,\ell}^2N_{\ell}\int_0^1\frac{\epsilon^{2+\sqrt{1+4\alpha_{\ell}^2}}}{\exp\left( \frac{\beta \delta r f''_{ext}(r_s)\alpha_{\ell}\epsilon}{2} \right)-1}d\epsilon\\
    &=\frac{N_b\delta r^2 \left[f''_{ext}(r_s)\right]^2}{8\pi} \sum_{\ell=1}^\infty N_{\ell} \int_0^1 \frac{\alpha_\ell^2\epsilon^{2+\sqrt{1+4\alpha_{\ell}^2}}}{\exp\left( \frac{\beta \delta r f''_{ext}(r_s)\alpha_{\ell}\epsilon}{2} \right)-1}d\epsilon. \\
    \end{split}
\end{equation}
Note that the integrand will generally have a local maximum. In particular, as long as $\beta \delta r f''_{ext}(r_s)$ is sufficiently large (this is true when we are sufficiently close to extremality) we are guaranteed to have a sharply peaked local maximum within the interval of integration. This means that we can easily extend the range of integration from $\epsilon \in (0,1)$ to $\epsilon \in (0,\infty)$ and still have a good estimate on the value of the integral. Such an integral can be done in full generality shown below:
\begin{equation}
\begin{split}
\label{firstnearextintapprox}
    &\int_0^1\frac{\alpha_\ell^2\epsilon^{2+\sqrt{1+4\alpha_{\ell}^2}}}{\exp\left( \frac{\beta \delta r f''_{ext}(r_s)\alpha_{\ell}\epsilon}{2} \right)-1}d\epsilon \approx \int_0^\infty\frac{\alpha_\ell^2\epsilon^{2+\sqrt{1+4\alpha_{\ell}^2}}}{\exp\left( \frac{\beta \delta r f''_{ext}(r_s)\alpha_{\ell}\epsilon}{2} \right)-1}d\epsilon\\
    &=\alpha_\ell^2\left(\frac{\beta \delta r f''_{ext}(r_s)\alpha_{\ell}}{2}\right)^{-3-\sqrt{1+4\alpha_\ell^2}}\Gamma\left( 3+\sqrt{1+4\alpha_{\ell}^2} \right)Li_{3+\sqrt{1+4\alpha_\ell^2}}\left(1\right).\\
    \end{split}
\end{equation}

Now we will deal with the second term(s) in Eq. (\ref{NearExtEvapRateInteg}) which describes modes with $\omega>\omega_{min,\ell}$ the terms read:
\begin{equation}
    \begin{split}
        &\frac{N_b}{2\pi}\sum_{\ell=1}^\infty \omega^2_{min,\ell} N_{\ell} \int_1^\infty\frac{\epsilon}{e^{\beta\omega_{min,\ell}\epsilon}-1}d\epsilon\\
        &=\frac{N_b}{2\pi\beta^2}\sum_{\ell=1}^{\infty}N_{\ell}\left[Li_2\left(e^{-\beta\omega_{min,\ell}}\right)-\beta\omega_{min,\ell}\ln\left(1-e^{-\beta\omega_{min,\ell}}\right)\right].\\
    \end{split}
\end{equation}
Combing these results give us the series representation of the evaporation rate given in Eq. (\ref{SeriesRepofEvapRateNearExt}).

\section{The Contribution to Evaporation Rate of Tunneling vs Non-Tunneling modes in Near Extremal Regime}
\label{TunVSnonTun}
In this appendix we will discuss the relative size between the terms $S_{tun,\ell}$ and $S_{ntun,\ell}$ which are used to define the evaporation rate of a near extremal AdS RN black hole. The goal is to estimate the following ratio:
\begin{equation}
\begin{split}
&\frac{S_{ntun,\ell}}{S_{tun,\ell}}=\frac{\left[Li_2\left(e^{-x_{min,\ell}}\right)-x_{min,\ell}\ln\left(1-e^{-x_{min,\ell}}\right)\right]x_{min,\ell}^{1+\sqrt{1+4\alpha^2_\ell}}}{\Gamma\left( 3+\sqrt{1+4\alpha_\ell^2} \right)Li_{3+\sqrt{1+4\alpha_\ell^2}}(1)}\\
&\approx \frac{x_{min,\ell}^{2+\sqrt{1+4\alpha^2_\ell}}e^{-x_{min,\ell}}}{\Gamma\left(3+\sqrt{1+4\alpha_\ell^2}\right)Li_{3+\sqrt{1+4\alpha_\ell^2}}(1)},\\
\end{split}
\end{equation}
where in the last line we used $x_{min,\ell}\gg 1$ since we are in the near extremal regime. For a fixed $\ell$ we can see that the ratio $S_{ntun,\ell}/S_{tun,\ell}\ll 1$ due to the exponential suppression and it follows that $S_{ntun,\ell}\ll S_{tun,\ell}$ when we are sufficiently close to the extremal regime. This is why we use the approximation in Eq. (\ref{OnlyTunEvaprateSer}) and ignore the modes that do not tunnel.

\section{Analysis of the Evaporation Rates of Near Extremal Very Large AdS RN Black Holes}
\label{NumericalAnalysisNearExtrBH}

We discuss approximating the integrand in Eq. (\ref{integrallargeextBH}) which is given by:
\begin{equation}
\begin{split}
    &I_{\ell}=\left(\frac{\alpha_\ell r_s^2f''_{ext}(r_s)}{(d-2)\sqrt{1+\frac{2r_s^2f''_{ext}(r_s)\alpha_\ell^2}{(d-2)^2}}}\right)\left(\frac{d-2}{2}\left[-1+\sqrt{1+\frac{2r_s^2f''_{ext}(r_s)\alpha_\ell^2}{(d-2)^2}}\right]\right)^{d-2}\\
    & \times \left[ \frac{\Gamma\left(3+\sqrt{1+4\alpha_{\ell}^2}\right)Li_{3+\sqrt{1+4\alpha_{\ell}^2}}(1)}{\left(\frac{\beta\delta r f''_{ext}(r_s)\alpha_\ell}{2}\right)^{1+\sqrt{1+4\alpha_\ell^2}}} \right]. \\
    \end{split}
\end{equation}
When $\alpha_\ell$ is close to zero one will find that the integrand initially grows. This growth will eventually slow down and stop when $\alpha_\ell$ is sufficiently large and the integrand will decay. A conventional leading order expansion of the integrand in the small or large $\alpha_{\ell}$ regime will not be able to capture this behavior. We make the following approximations, the product of the first two terms is approximated in the large $\alpha_{\ell}$ regime to give:
\begin{equation}
\begin{split}
& \left(\frac{\alpha_\ell r_s^2f''_{ext}(r_s)}{(d-2)\sqrt{1+\frac{2r_s^2f''_{ext}(r_s)\alpha_\ell^2}{(d-2)^2}}}\right)\left(\frac{d-2}{2}\left[-1+\sqrt{1+\frac{2r_s^2f''_{ext}(r_s)\alpha_\ell^2}{(d-2)^2}}\right]\right)^{d-2}\\
    & \approx \left(\frac{r_s^2f_{ext}''(r_s)}{2}\right)^{\frac{d-1}{2}}\alpha_\ell^{d-2}.  \\
    \end{split}
\end{equation}
Even though this approximation is more accurate for larger $\alpha_{\ell}$ one can plot the approximation and compare to the exact function and find reasonable agreement at small values of $\alpha_{\ell}$. 

\begin{figure}[h!]
\centering
\includegraphics[width=150mm]{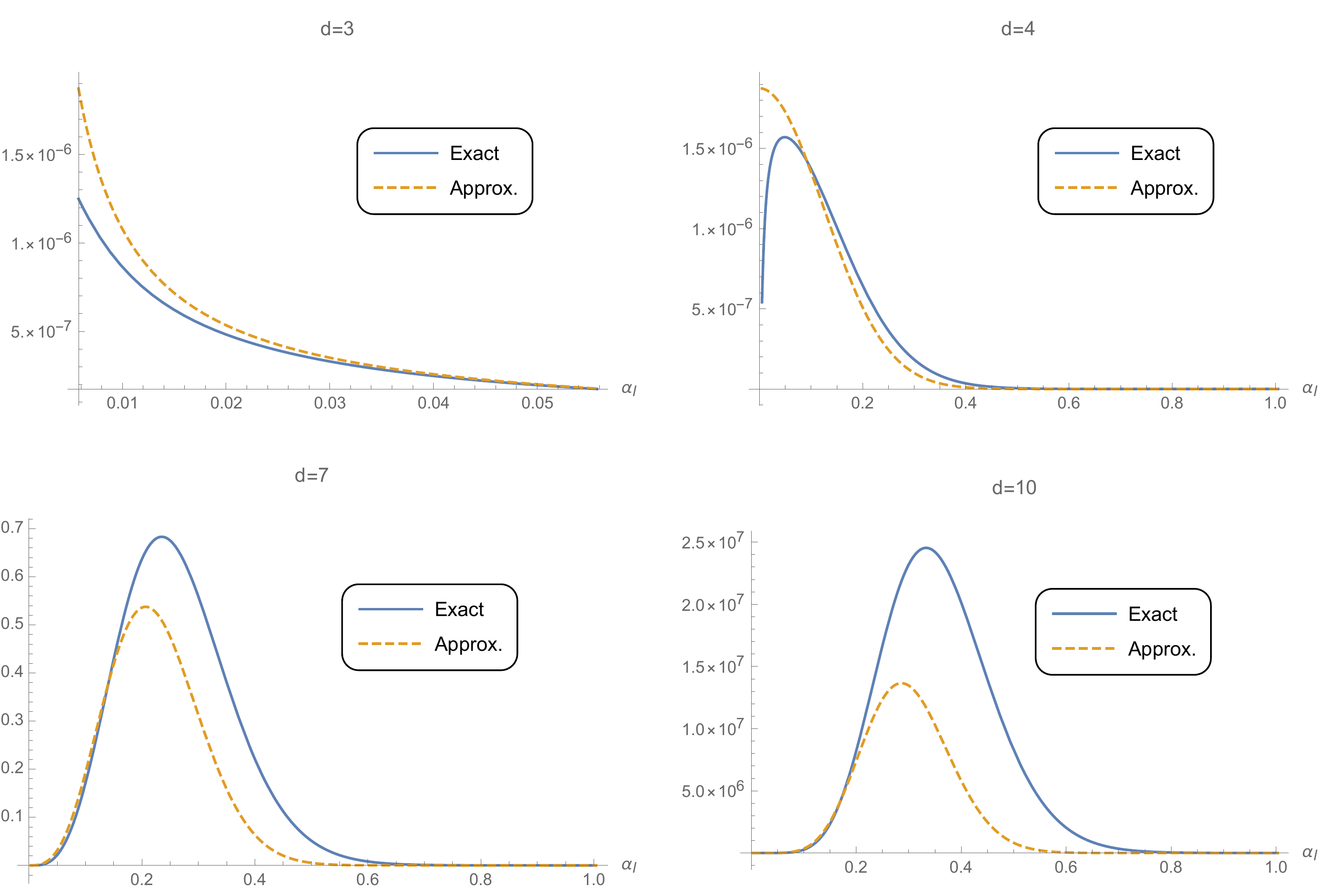}
\caption{$\frac{\beta\delta r}{r_s^2}=100$ and $r_s/L=100$.  \label{IntegrandPlotsbdel100r100} }
\end{figure}

We approximate the terms in the second line with:
\begin{equation}
    \left[ \frac{\Gamma\left(3+\sqrt{1+4\alpha_{\ell}^2}\right)Li_{3+\sqrt{1+4\alpha_{\ell}^2}}(1)}{\left(\frac{\beta\delta r f''_{ext}(r_s)\alpha_\ell}{2}\right)^{1+\sqrt{1+4\alpha_\ell^2}}} \right]\approx \frac{\pi^4}{15}\left(\frac{\beta \delta r f''_{ext}(r_s)}{2}\right)^{-2-2\alpha_{\ell}^2}\alpha_{\ell}^{-2}.
\end{equation}
Combining these gives:
\begin{equation}
    I_{\ell}\approx \frac{\pi^4}{15}\left(\frac{r_s^2f''_{ext}(r_s)}{2}\right)^{\frac{d-1}{2}}\left(\frac{\beta\delta r f''_{ext}(r_s)}{2}\right)^{-2-2\alpha_\ell^2}\alpha_{\ell}^{d-4}.
\end{equation}
To get a sense of how the approximation compares to the full function we make various plots shown in Figure \ref{IntegrandPlotsbdel100r100} by fixing the values of $\frac{\beta\delta r}{r_s^2}$, $\frac{r_s}{L}$, and $d$. We can see that the approximation becomes worse as we increase $d$. However, even for larger values of $d$ doing the integral of the approximated integrand will give a result that is off by a order one pre-factor from the exact result. 
The approximation improves in the limit when $r_s/L\to \infty$ as we can see in the following plots in Figure \ref{IntegrandPlotsbdel100r10to50} where we increase $r_s/L$ from $10^2$ to $10^{50}$: 

\begin{figure}[h!]
\centering
\includegraphics[width=150mm]{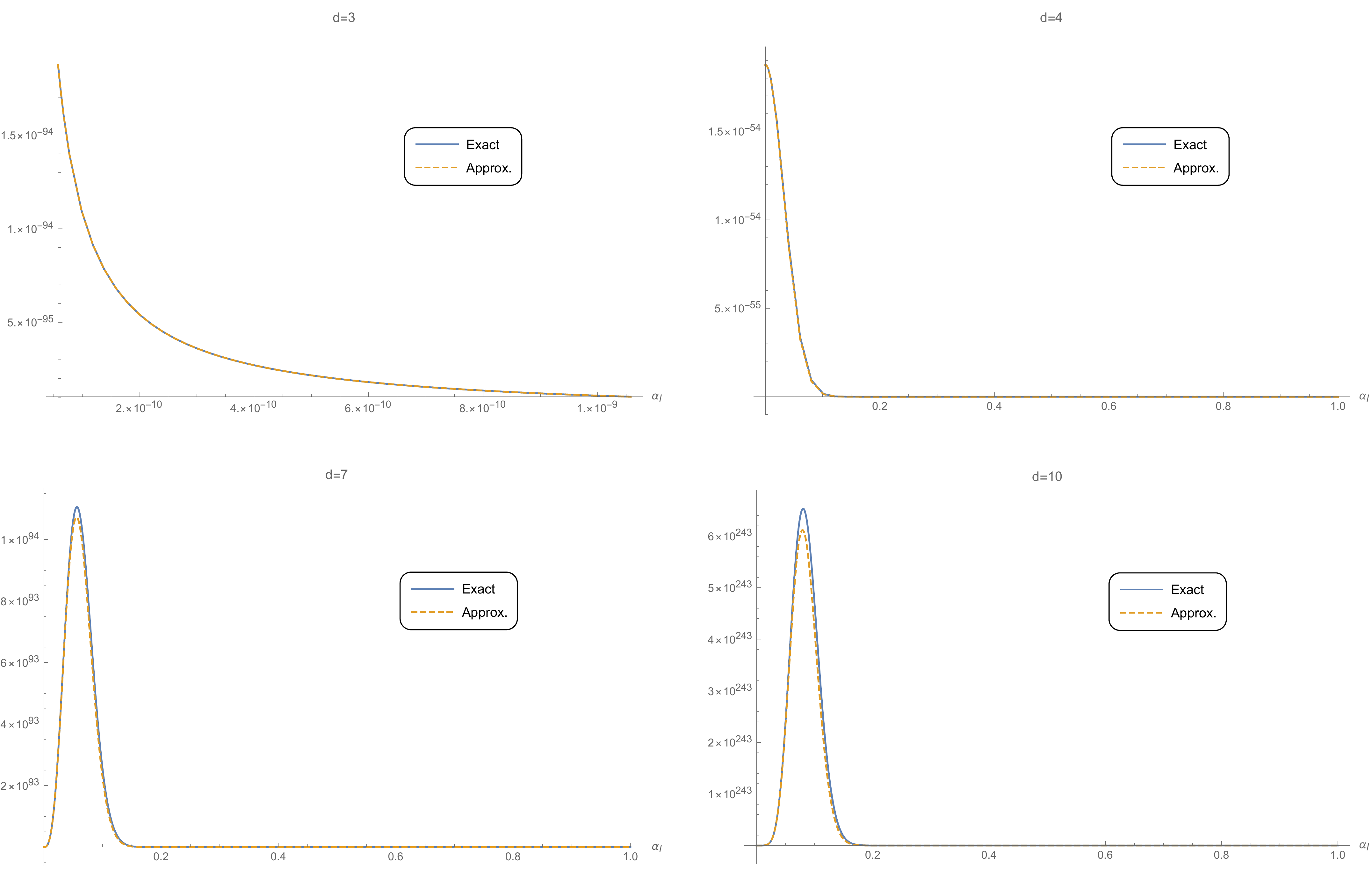}
\caption{$\frac{\beta\delta r}{r_s^2}=100$ and $r_s/L=10^{50}$.  \label{IntegrandPlotsbdel100r10to50} }
\end{figure}

\section{Treating the $\ell=0$ Mode in the Near Extremal Regime}
\label{elleq0mode}
Using the generalized greybody factor given by Eq. (\ref{GGBFleeq0}) the contribution to the evaporation rate of a near extremal AdS RN black hole is given by computing the following integrals:
\begin{equation}
\begin{split}
    &\frac{dM}{dt}\bigg\vert_{\ell=0}=\frac{N_b}{2\pi}\left[ \int_0^{\omega_{min,0}} \frac{\omega}{e^{\beta\omega}-1}\left(\frac{\omega}{\omega_{min,0}}\right)^{4/3}d\omega+\int_{\omega_{min,0}}^\infty\frac{\omega}{e^{\beta\omega}-1}d\omega \right]\\
    &=\frac{N_b\omega_{min,0}^2}{2\pi}\left[ \int_0^1\frac{\epsilon^{10/3}}{e^{\beta\omega_{min,0}\epsilon}-1}d\epsilon+\int_1^\infty\frac{\epsilon}{e^{\beta\omega_{min,0}\epsilon}-1}d\epsilon \right].\\
    \end{split}
\end{equation}
Similar to the $\ell\geq 1$ modes, when $\beta\omega_{min,0}\gg 1$ we can extend the integration limits over the interval $(0,1)$ to $(0,\infty)$ without changing the value of the result. When we do this we will obtain the following evaporation rate from the $\ell=0$ channel:
\begin{equation}
    \frac{dM}{dt}\bigg\vert_{\ell=0}=\frac{N_b}{2\pi\beta^2}\left[\frac{\Gamma\left(\frac{13}{3}\right)\zeta\left(\frac{13}{3}\right)}{\left(\beta\omega_{min,0}\right)^{7/3}}+Li_2\left(e^{-\beta\omega_{min,0}}\right)-\beta\omega_{min,0}\ln\left(1-e^{\beta\omega_{min,0}}\right)\right],
\end{equation}
where $\Gamma\left( 13/3 \right)\zeta(13/3)\approx 9.84$. Close to the extremal regime the primary contribution comes from the first term just like for modes $\ell\geq 1$. So we have:
\begin{equation}
\label{leq0evaprate}
    \frac{dM}{dt}\bigg\vert_{\ell=0}\approx \frac{N_b}{2\pi\beta^2}\Gamma\left(13/3\right)\zeta\left(13/3\right)\left[\left(\frac{(d-1)\delta r}{r_s}\right)^{1/2}\frac{\beta\delta r f''_{ext}(r_s)}{2}\right]^{-7/3}.
\end{equation}
Now that we have the $\ell=0$ contribution to the evaporation rate it is interesting to ask how large of an effect it has on the evaporation rate if we include it. 

For large AdS black holes we estimated the evaporation rate contribution from modes with $\ell\geq 1$. We found that the evaporation rate went as $\beta^{-4}$ (perhaps with some $\ln(\beta)$ dependence which we can ignore if $\beta$ is sufficiently large.). If we look at the $\beta$ dependence of the $\ell=0$ contribution to the evaporation rate is it goes as $\beta^{-13/3}$. In the extremal limit when $\beta\to \infty$ the $\ell=0$ mode's contribution to the evaporation rate will decay more quickly than the total contribution of the modes $\ell\geq 1$. So sufficiently close to the extremal regime for very large AdS black holes we can effectively ignore the contribution of the the $\ell=0$ mode since it will be a sub-leading correction. This is also what happened in the case of AdS Schwarzschild black holes (assuming we placed the screen sufficiently close to the horizon).   

For very small AdS black holes (or asymptotically flat black holes) we can consider the ratio between the $\ell=1$ mode evaporation rate (given by Eq. (\ref{smallAdSbhNearExtEvapRate})) and the $\ell=0$ evaporation rate (given by Eq. (\ref{leq0evaprate})) which is given by:

\begin{equation}
\begin{split}
    &\frac{\frac{dM}{dt} \bigg\vert_{\ell=0}}{\frac{dM}{dt}\bigg\vert_{\ell=1}}=\frac{\Gamma\left( 13/3 \right)\zeta\left( 13/3 \right)\left[ (d-1)(d-2)^2 \right]^{\frac{d-1}{d-2}}}{\left[(d-1)^{1/2}(d-2)^2\right]^{7/3}\Gamma\left( \frac{2(2d-3)}{d-2} \right)Li_{\frac{2(2d-3)}{d-2}}(1)}\left(\frac{r_s}{\delta r}\right)^{7/4}\left(\frac{r_s^2}{\beta\delta r}\right)^{\frac{d-8}{3(d-2)}}\\
    &\sim \beta^{-\frac{d-8}{3(d-2)}}.\\
    \end{split}
\end{equation}
We can see that if $d<8$ then $\ell=0$ mode will dominate sufficiently close to the extremal regime and if $d>8$ then the $\ell=1$ mode will dominate sufficiently close to the extremal regime. Lets consider what happens when $d<8$. For $\ell=0$ mode we require $\beta\omega_{min,0}\gg 1 \Rightarrow \frac{\delta r}{r_s}\gg \left(\frac{r_s}{\beta}\right)^{2/3}$. We define $\Lambda$ such that $\frac{\delta r}{r_s}=\Lambda \left(\frac{r_s}{\beta}\right)^{2/3}$ then the evaporation rate can be expressed as:
\begin{equation}
    \frac{dM}{dt}\sim \frac{N_b}{\beta^2\Lambda^{7/2}}.
\end{equation}
Using this result we can express the information re-emergence time as:
\begin{equation}
    t_{emerge}\sim \frac{\beta}{2\pi}\ln\left(\Lambda^{7/2}\frac{r_s^{d-1}}{\ell_{p}^{d-1}}\right).
\end{equation}
We fix the $\beta$ dependence of $\Lambda$ by fixing the proper distance (which we denote as $l_{prop}$) between the screen and horizon. This means $\Lambda\sim \left(\frac{r_s}{\beta}\right)^{1/3}\left(\frac{l_{prop}}{r_s}\right)^2$. Then we find:
\begin{equation}
    t_{emerge}\sim \frac{\beta}{2\pi}\left[\ln\left( \left(\frac{r_s}{\beta}\right)^{7/6}\frac{r_s^{d-1}}{\ell_p^{d-1}N_b} \right)+2\ln\left(\frac{l_{prop}}{r_s}\right)\right].
\end{equation}
This shows that the even when $\ell=0$ dominates we still get results for the information re-emergence time that are comparable to the results we obtained in cases where the $\ell=1$ mode was dominant. The main change is the power that the combination $r_s/\beta$ comes with, which is $7/6$ rather than $1$. As before, the sub-leading log term contains information about how far from the horizon the screen is placed. So we again get results consistent with known literature on the scrambling time of near extremal black holes. 

\section{Derivation of the Stress Energy Tensor of the Screen}
\label{DerivationOfShellEnergy}
In Section \ref{RigourousApproachtoGBF} we suggested that generalized greybody factors discussed in this paper can be understood in terms of a transmission coefficient for an effective potential given in Eq. (\ref{ScreenEffectivePotential}). We stated that such a potential would be obtained by cutting off the geometry of the AdS black hole where the screen would be, we would then glue an exterior space which acts as a storage system for the radiation. By requiring this gluing to satisfy the Einstein field equations with some matter distribution, then there will generally be a singular matter distribution at the interface where the gluing occurs. In our setup the singular matter will lie on a spherical shell where the screen would be. The problem of finding the stress energy tensor of such a shell is a well studied problem whose solution is stated in terms of Israel junction conditions \cite{Israel1966} (see \cite{poisson_2004} for a review). The starting point is to write down the metrics both inside and outside the shell. We will utilize a hyperspherical coordinate system $x^{\mu}=(t,r,\phi^1,...,\phi^{d-1})$. This coordinate sytem will be used both inside and outside the shell. In these coordinates the shell is at a fixed at $r=r_0=const$. The metric inside the shell will be given by:

\begin{equation}
    g^-_{\mu\nu}=-f_-(r)\delta^t_\mu\delta^t_\nu+f_-(r)^{-1}\delta^r_\mu\delta^r_\nu+r^2g^{\Omega}_{IJ}\delta^I_\mu\delta^J_\nu,
\end{equation}
where $I,J=1,2,...,d-1$ are angular indices and $g^{\Omega}_{IJ}$ is the (diagonal) metric on a $d-1$ - unit sphere. The ``$-$'' sub-indices and super-indices indicate that we are dealing with tensors inside the shell $(r< r_0)$. Analogously, we take the metric outside the shell to be:

\begin{equation}
\begin{split}
    &g^+_{\mu\nu}=-\Delta(r_0)f_+(r)\delta^t_\mu\delta^t_\nu+f_+(r)^{-1}\delta^r_\mu\delta^r_\nu+r^2g^{\Omega}_{IJ}\delta^I_\mu\delta^J_\nu\\
    &\Delta(r_0)=\frac{f_-(r_0)}{f_+(r_0)},\\
    \end{split}
\end{equation}
 where the ``$+$'' sub-indices and super-indices indicate that we are dealing with tensors outside the shell $(r> r_0)$. The additional time lapse constant $\Delta(r_0)$ is introduced so that the induced metric on both sides of the shell is the same in ``natural'' induced coordinates $y^{a}=(t,\phi^1,...,\phi^{d-1})$. It is given as:
 
 \begin{equation}
     h_{ab}=\frac{\partial x^{\mu}}{\partial y^a}\frac{\partial x^{\nu}}{\partial y^b}g^{\pm}_{\mu\nu}\bigg|_{r=r_0}=\delta^\mu_a\delta^\nu_bg_{\mu\nu}^{\pm}=-f(r_0)\delta^t_a\delta^t_b+r_0^2g^{\Omega}_{IJ}\delta^I_a\delta^J_b.
 \end{equation}
The stress energy tensor of the shell denoted $S_{ab}$ is related to the discontinuity in the extrinsic curvature tensor and its trace on either side of the $r=r_0$ hypersurface. More specifically, we have\footnote{In the formula below we assume that hypersurface is timelike.}:

\begin{equation}
\label{StressEnergyofShell}
    S_{ab}=-\frac{1}{8\pi}\left([K_{ab}]-[K]h_{ab}\right),
\end{equation}
above the notation $[T]$ for any tensor $T$ is defined as:

\begin{equation}
    [T]=\lim_{r\to r_0}T^+-\lim_{r\to r_0}T^-.
\end{equation}
So we need to calculate the extrinsic curvature on either side of the hypersurface, which is defined in terms of the covariant derivative of the normalized unit vector to the timelike hypersurface $r=r_0$:
\begin{equation}
    K^{\pm}_{ab}=\delta^{\mu}_a\delta^{\nu}_b\nabla^{\pm}_\mu n^{\pm}_{\nu}.
\end{equation}
Here, $\nabla^{\pm}_{\mu}$ is the covariant derivative with respect to the metrics, $g_{\mu\nu}^{\pm}$, on either side of the shell. The trace is simply given by:

\begin{equation}
    K^\pm=h^{ab}K^{\pm}_{ab}.
\end{equation}
The normal vector to a constant $r$ hypersurface outside the shell is:

\begin{equation}
    n_\mu^+=f_+(r)^{-1/2}\delta^r_\mu.
\end{equation}
The normal vector to a constant $r$ hypersurface inside the shell is:
\begin{equation}
  n_\mu^-=f_-(r)^{-1/2}\delta^r_\mu.
\end{equation}
Using these expressions we will find that:

\begin{equation}
\begin{split}
    &K^+_{ab}=-\frac{1}{2}f_-(r_0)f_+(r_0)^{-1/2}f'_+(r_0)\delta^t_a\delta^t_b+r_0f_+(r_0)^{1/2}g^{\Omega}_{IJ}\delta^I_a\delta^J_b\\
    &K^-_{ab}=-\frac{1}{2}f_-(r_0)^{1/2}f'_-(r_0)\delta^t_a\delta^t_b+r_0f_-(r_0)^{1/2}g^\Omega_{IJ}\delta^I_a\delta^J_b \\
    &K^+=\frac{1}{2}f_+(r_0)^{-1/2}f_+'(r_0)+(d-1)r_0^{-1}f_+(r_0)^{1/2} \\
    &K^-=\frac{1}{2}f_-(r_0)^{-1/2}f'_-(r_0)+(d-1)r_0^{-1}f_-(r_0)^{1/2}.\\
    \end{split}
\end{equation}
Using these expressions and plugging into Eq. (\ref{StressEnergyofShell}) gives:

\begin{equation}
\begin{split}
    &16\pi S_{ab}=-\frac{2(d-1)f_-(r_0)\left( f_+(r_0)^{1/2}-f_-(r_0)^{1/2} \right)}{r_0}\delta^t_a\delta^t_b\\
    &+\left[ 2(d-2)\left( f_+(r_0)^{1/2}-f_-(r_0)^{1/2} \right)+\frac{r_0f'_+(r_0)}{f_+(r_0)^{1/2}}-\frac{r_0f'_-(r_0)}{f_-(r_0)^{1/2}} \right]r_0g^{\Omega}_{IJ}\delta^I_a \delta^J_b.\\
    \end{split}
\end{equation}
It is convenient to define the following basis on the shell:

\begin{equation}
\begin{split}
    &\hat{e}^a_t=\frac{\delta^a_t}{\sqrt{f_-(r_0)}}\\
    &\hat{e}^a_I=\frac{\sqrt{g^{II}_{\Omega}}}{r_0}\delta^a_I,\\
    \end{split}
\end{equation}
which allows us to write the inverse induced metric as:

\begin{equation}
    h^{ab}=\eta^{cd}\hat{e}^a_c\hat{e}^b_d=-\hat{e}^a_t\hat{e}^b_t+\sum_{I=1}^{d-1}\hat{e}^a_I\hat{e}^b_I.
\end{equation}
Using this basis we can see the stress energy tensor of the shell is that of a $d$ - dimensional perfect fluid given by:

\begin{equation}
\begin{split}
\label{StressEnergy2}
    S^{ab}&=\rho\hat{e}^a_t\hat{e}^b_t+p\sum_I\hat{e}^a_I\hat{e}^b_I=\left(\rho+p\right)\hat{e}^a_t\hat{e}^b_t+ph^{ab}\\
    &\rho=\frac{(d-1)\left( f_-(r_0)^{1/2}-f_+(r_0)^{1/2} \right)}{8\pi r_0}\\
    &p=\frac{1}{16\pi r_0}\left[ 2(d-2)\left(f_+(r_0)^{1/2}-f_-(r_0)^{1/2}\right)+r_0\left(\frac{f'_+(r_0)}{f_+(r_0)^{1/2}}- \frac{f'_-(r_0)}{f_-(r_0)^{1/2}} \right) \right],\\
    \end{split}
\end{equation}
where $\rho$ is the energy density of the shell and $p$ is the principle pressure. This completes our derivation of the stress energy tensor of a shell that allows for the gluing two spherically symmetric static spacetimes along the interface $r=r_0$. This will be used in the discussion of energy conditions of the shell.

\section{Null Energy Condition of the Screen}
\label{ScreenNEC}
In Appendix \ref{DerivationOfShellEnergy}, we derived a solution to the Einstein equation which represented the gluing of two different spherically symmetric solutions to the Einstein equation along a timelike hypersurface $r=r_0$ where our ``absorptive'' screen would be placed\footnote{The reason for quotation marks is that the radiation is not actually absorbed by the screen, but rather leaks into the exterior flat or AdS space.}. To have a consistent patching it is required that there be a thin shell of matter with a stress energy tensor given by Eq. (\ref{StressEnergy2}). It is interesting to ask if such a shell will satisfy energy conditions. 

In particular, we are interested in the null energy condition (NEC). The NEC states that for any future directed null vector $k^{\mu}$ one has:

\begin{equation}
    T_{\mu\nu}k^\mu k^{\nu}\geq 0.
\end{equation}
If we restrict ourselves to null vectors with no radial component then NEC simply becomes:

\begin{equation}
    \rho+p\geq 0.
\end{equation}
On the other hand, considering a purely radial null vector is more subtle since the $rr$ component of the metric is discontinuous across the shell, and we should consider what happens on each side separately. The null vector will be given by: 

\begin{equation}
    k^\mu_{\pm}=c_{\pm}\left[ \delta^\mu_t +\left(\frac{-g_{tt}^{\pm}}{g_{rr}^{\pm}}\right)^{1/2}\delta^\mu_r \right].
\end{equation}
If the radial null vector is to be future directed then $c_{\pm}\geq 0$. Since the stress energy tensor of the shell has no radial component we see that the null energy condition for a radial null vector becomes:

\begin{equation}
    \rho\geq 0,
\end{equation}
which is to say that the matter on the shell has a positive energy density. Now let us consider interior metric to be that of a Schwarzschild AdS black hole:

\begin{equation}
    f_-(r)=1+\frac{r^2}{L_-^2}-\left(\frac{r_H}{r}\right)^{d-2}\left( 1+\frac{r_H^2}{L_-^2} \right).
\end{equation}
The exterior metric will be chosen to be that of pure AdS\footnote{The reader might be wondering why we choose pure AdS rather than flat space as we suggested in Section \ref{RigourousApproachtoGBF}. The reason is that we want to have a well defined holographic description of the exterior system where the radiation is stored. The flat space limit can be obtained by sending $L_+$ to infinity. The advantage of using pure AdS rather than flat space from the beginning is that we can control how much separation there is between the shell and the exterior conformal boundary, the larger $L_+$ is the further we push the conformal boundary away from the screen. }:
\begin{equation}
     f_+(r)=1+\frac{r^2}{L_+^2}.
\end{equation}
Before analyzing whether it is possible to have $\rho\geq 0$ we will consider what happens to the energy density of the screen as we approach the horizon and the conformal boundary. At the horizon the the energy density of the screen takes on a negative value given by:

\begin{equation}
\label{EnergyDensityAtHorizon}
    \rho(r_0=r_H)=-\frac{d-1}{8\pi r_H}\sqrt{1+\frac{r_H^2}{L_+^2}}.
\end{equation}
As the screen gets closer to the conformal boundary the energy density will saturate to the following constant:

\begin{equation}
\label{EnergyDensityAtBdry}
    \lim_{r_0\to \infty}\rho =\frac{d-1}{8\pi}\left(\frac{1}{L_-}-\frac{1}{L_+}\right).
\end{equation}
From this we see that the radial NEC is always violated at the horizon but if $L_-\leq L_+$ then the radial NEC is satisfied as the screen approaches the conformal boundary. Now we will discuss the constraint of the energy density being non-negative. It will read: 

\begin{equation}
    \sqrt{1+\frac{r_0^2}{L_-^2}-\left( \frac{r_H}{r_0} \right)^{d-2}\left( 1+\frac{r_H^2}{L_-^2} \right)}\geq \sqrt{1+\frac{r_0^2}{L_+^2}}.
\end{equation}
Under the assumption that $r_0>r_H$ we can square the expressions on both sides of the inequality to obtain the following simplified constraint: 

\begin{equation}
\label{RadialNEC}
    \frac{1}{L_-^2}-\frac{1}{L_+^2}-\frac{1}{r_H^2}\left( \frac{r_H}{r_0} \right)^d\left(1+\frac{r_H^2}{L_-^2}\right)\geq 0.
\end{equation}
We already know the radial NEC will be satisfied for a screen at the conformal boundary if $L_+\geq L_-$. Furthermore, we also know that if screen is placed arbitrarily close to the horizon the radial NEC will be violated. From these considerations there must be a critical radius where the screen will saturate the radial NEC and the energy density will vanish. This is easily found and given by:  

\begin{equation}
\label{CriticalPoint}
   r_c=r_H\left(\frac{1+\frac{L_-^2}{r_H^2}}{1-\frac{L_-^2}{L_+^2}}\right)^{\frac{1}{d}}.
\end{equation}

\begin{figure}[h!]
\centering
\includegraphics[width=150mm]{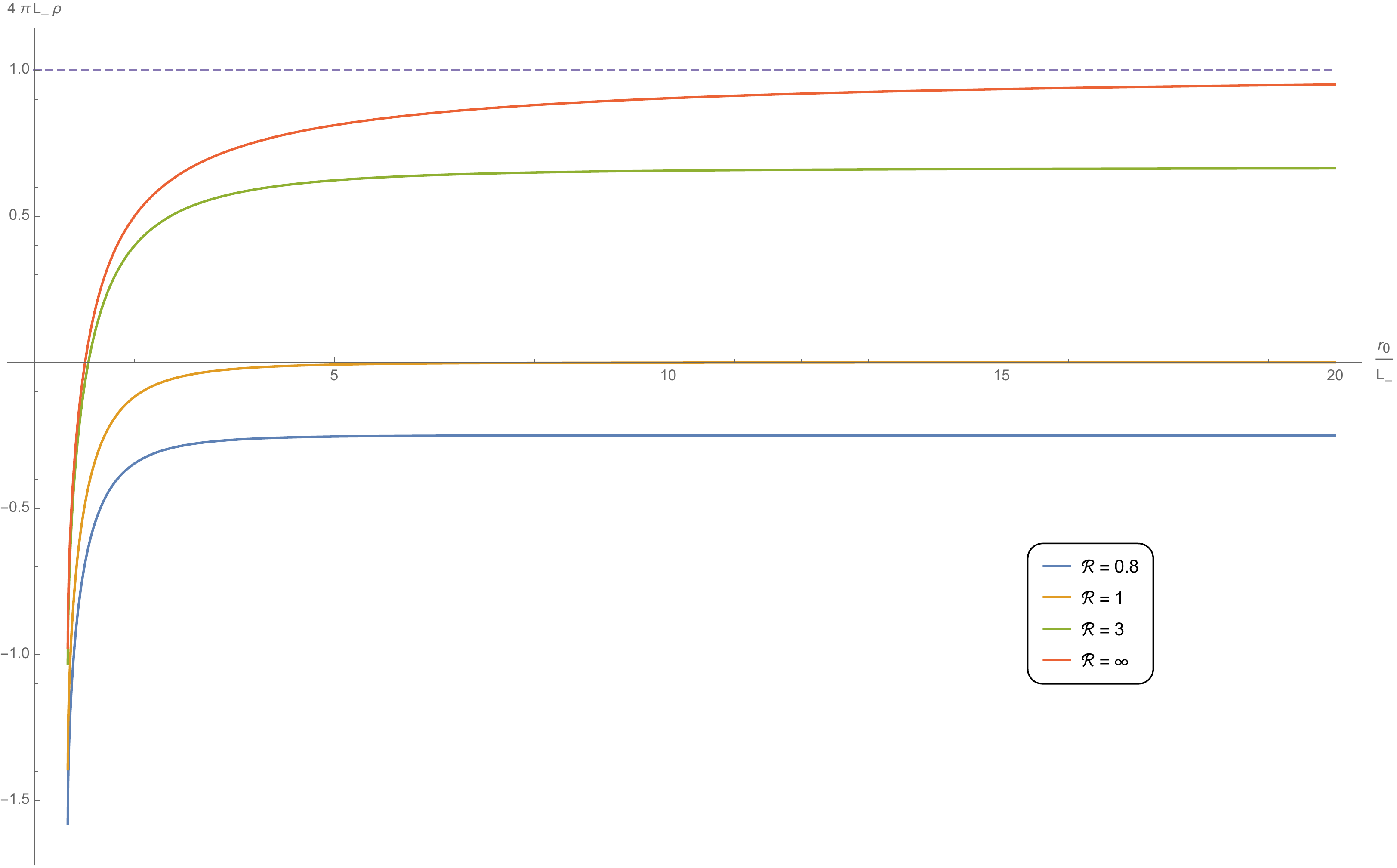}
\caption{Above is a plot of the energy density of the screen as a function of its placement for the case when $d=3$ and $r_H/L_-=1$. Each solid line is a plot of the energy density of the screen in units of the interior AdS radius, $L_-$ for different choices of the ratio $\mathcal{R}=L_+/L_-$. We can see that all the lines start at $r_0/L_-=1$ which is where the horizon of the black hole is. At $r_0/L_-=1$ the energy density given by Eq. (\ref{EnergyDensityAtHorizon}) and will be negative. All the lines then increase monotonically and will saturate to a value given by Eq. (\ref{EnergyDensityAtBdry}) at infinity. For cases when $\mathcal{R}<1$ the energy density at the conformal boundary will saturate to a negative value. When $\mathcal{R}=1$ the energy density is always negative and saturates to zero at the conformal boundary. When $\mathcal{R}>1$ the energy density is positive if $r_0>r_c$ where $r_c$ is given by Eq. (\ref{CriticalPoint}). The red curve corresponding to the limit when $\mathcal{R}=\infty$ represents the case when we patch a flat exterior metric at the screen interface and the dotted line is the value the energy density will saturate to at infinity. The main features of the energy density as illustrated in this plot remain intact if we consider higher dimensions and different values of $r_H/L_-$. \label{EnergyDensityPlot} }
\end{figure}

It is also interesting to consider how the energy density of the screen changes as we move the screen closer to the conformal boundary by considering $d\rho/dr_0\geq 0$ for any radial coordinate outside the horizon. The expression for the derivative is given by:

\begin{equation}
\begin{split}
    \frac{d\rho}{dr_0}&=\frac{d-1}{8\pi r_0^2}\left[\frac{r_0f_-'(r_0)}{2\sqrt{f_-(r_0)}}-\frac{r_0f_+'(r_0)}{2\sqrt{f_+(r_0)}}-\sqrt{f_-(r_0)}+\sqrt{f_+(r_0)} \right]\\
    &=\frac{(d-1)\xi(r_0)}{8\pi r_0^2\sqrt{f_-(r_0)f_+(r_0)}}\\
    &\xi(r_0)= \sqrt{f_+(r_0)}\left( \frac{r_0f'_-(r_0)}{2}-f_-(r_0) \right) +\sqrt{f_-(r_0)}\left(f_+(r_0)-\frac{r_0f_+'(r_0)}{2}\right).\\
    \end{split}
\end{equation}
The sign of the derivative depends on $\xi(r_0)$. By plugging in the expressions for $f_+(r_0)$ and $f_-(r_0)$ we will find that:

\begin{equation}
\begin{split}
    &\xi(r_0)=\sqrt{1+\frac{r_0^2}{L_+^2}}\left[ -1+\frac{d}{2}\left(\frac{r_H}{r_0}\right)^{d-2}\left(1+\frac{r_H^2}{L_-^2}\right) \right]+\sqrt{1+\frac{r_0^2}{L_-^2}-\left(\frac{r_H}{r_0}\right)^{d-2}\left(1+\frac{r_H^2}{L_-^2}\right)}\\
    & > \sqrt{1+\frac{r_0^2}{L_-^2}-\left(\frac{r_H}{r_0}\right)^{d-2}\left(1+\frac{r_H^2}{L_-^2}\right)} - \sqrt{1+\frac{r_0^2}{L_+^2}}=\frac{8\pi r_0\rho}{d-1}.\\
    \end{split}
\end{equation}
We have a strict inequality since $r_0<\infty$ (saturation occurs in limit as $r_0\to \infty$):
\begin{equation}
    \frac{d\rho}{dr_0} > \frac{\rho}{r_0\sqrt{f_+(r_0)f_-(r_0)}}.
\end{equation}
This implies that at any point where the radial NEC is satisfied the energy density must increase within a neighborhood of that point. This is enough to show that for $r_0 \geq r_c$ the energy density must strictly increase. In Figure \ref{EnergyDensityPlot} we plot of the energy density of the screen to illustrate the monotone increase of energy density.

Now that we have explored when the NEC is violated for radial null vectors we can move on to understanding the NEC for tangent null vectors (i.e. null vectors with no radial component). In this case we must understand the condition $\rho+p\geq 0$. Before doing this lets consider what happens to this combination as we approach the horizon and as we approach infinity. As we approach the horizon we have:

\begin{equation}
\label{TangNECHor}
    \lim_{r_0 \to r_H}\left(\rho+p\right)=-\infty.
\end{equation}
When we take the screen to infinity it can be shown that $\rho+p$ goes to zero with the following leading order behaviour:
\begin{equation}
\label{TangNECBdry}
    \rho+p=\frac{L_--L_+}{8\pi r_0^2}+\mathcal{O}\left(\frac{1}{r_0^4}\right).
\end{equation}
This means that if $L_+> L_-$ then for sufficiently large $r_0$ the sum of the energy density and pressure is negative. If $L_+<L_-$ the for sufficiently large $r_0$ the sum of the energy density and pressure is positive. Similar to the radial NEC, we see that there is a violation of the tangent NEC close to the horizon and a saturation at infinity. The divergent violation at the horizon comes from the pressure given by Eq. (\ref{StressEnergy2}) due to the fact that $f_-(r_H)=0$. Now that we understand what happens close to the horizon and infinity we will consider the constraint more generally. In terms of $f_+$ and $f_-$, it is given by the following inequality:

\begin{equation}
\label{TangentialNEC}
   \rho+p=\frac{1}{16\pi r_0}\left[2\left( f_-(r_0)^{1/2}-f_+(r_0)^{1/2} \right)+r_0\left(\frac{f_+'(r_0)}{f_+(r_0)^{1/2}}-\frac{f_-'(r_0)}{f_-(r_0)^{1/2}}\right)\right]\geq 0. 
\end{equation}

It is difficult to make further progress analytically like we did for understanding the radial NEC. Therefore, we will resort to making plots for $\rho+p$ in Eq. (\ref{TangentialNEC}) and make some general comments.

\begin{figure}[h!]
\centering
\includegraphics[width=150mm]{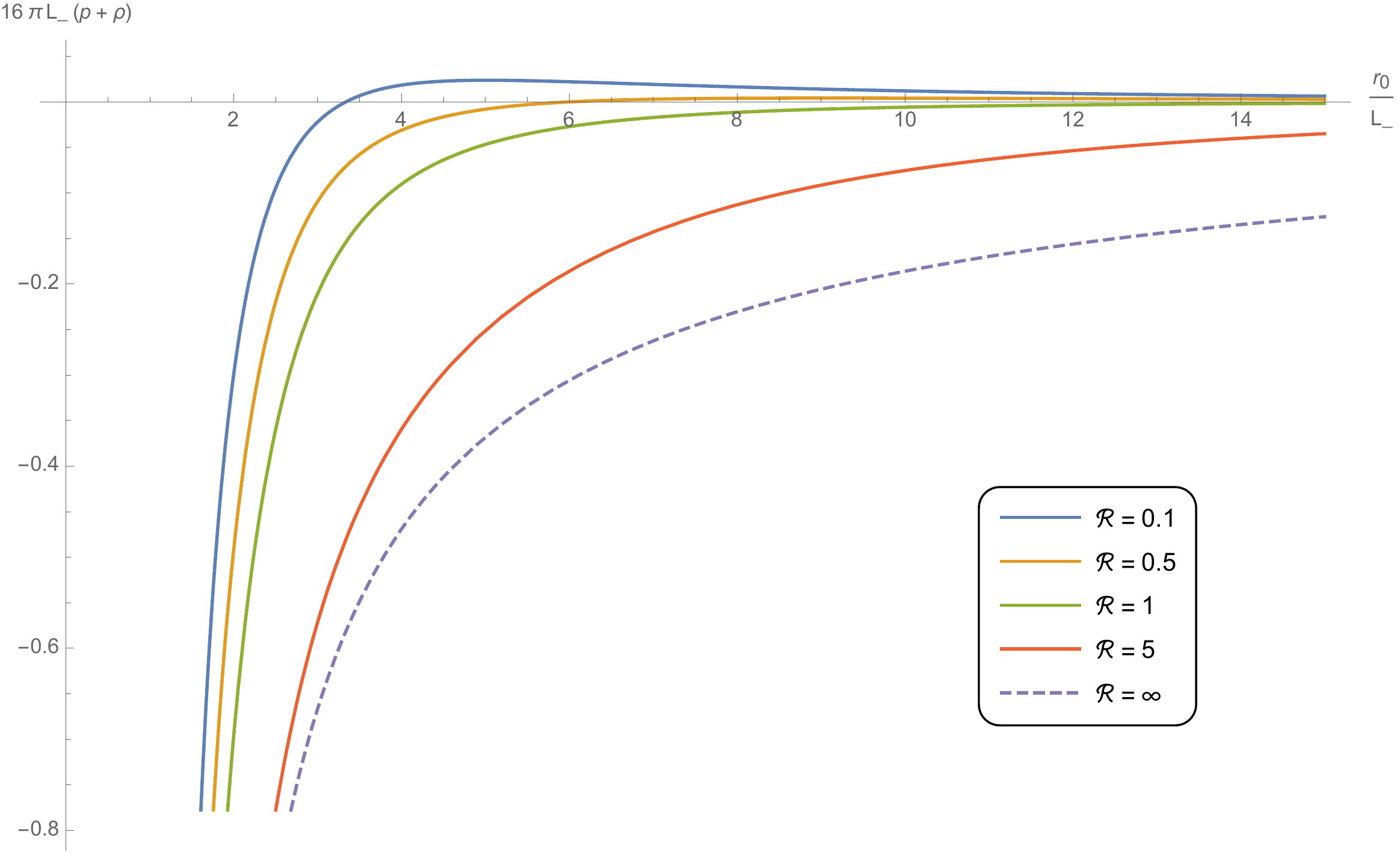}
\caption{Above is a plot of the sum of the energy density and pressure (i.e. $\rho+p$) of the screen as a function of its radial placement for the case when $d=3$ and $r_H/L_-=1$. Each solid line is a plot of the energy density of the screen in units of the interior AdS radius, $L_-$, for different choices of the ratio $\mathcal{R}=L_+/L_-$. For each line there is a divergence at $r_0/L_-=1$ where the horizon of the black hole is. All the lines in the graph will saturate to a value of zero at infinity, however the way by which this is achieved is different depending on the value of $\mathcal{R}$. Lines with $\mathcal{R}<1$ (blue and yellow line) actually cross the $x$-axis and then decrease and saturate to zero. Lines with $\mathcal{R}\geq 1$ (green, red, and dotted lines) stay below the $x$-axis and saturate to zero at infinity. This is consistent with the simple expressions we found in  Eq. (\ref{TangNECHor}) and Eq. (\ref{TangNECBdry}). The main features discussed remain intact in higher dimensions and for more general choices of $r_H/L_-$.  \label{TangentialNECPlot} }
\end{figure}

From Figure \ref{TangentialNECPlot} we can see that if $L_+\geq L_-$ then the tangential NEC is violated for all $r_0\in(r_H,\infty)$.

\bibliography{Ref.bib}

\providecommand{\href}[2]{#2}\begingroup\raggedright\begin{thebibliography}{10}

\bibitem{Maldacena:1997re}
J.~M. Maldacena, \emph{{The Large N limit of superconformal field theories and
  supergravity}}, \href{https://doi.org/10.1023/A:1026654312961,
  10.4310/ATMP.1998.v2.n2.a1}{\emph{Int. J. Theor. Phys.} {\bfseries 38} (1999)
  1113} [\href{https://arxiv.org/abs/hep-th/9711200}{{\ttfamily
  hep-th/9711200}}].

\bibitem{Ramallo:2013bua}
A.~V. Ramallo, \emph{{Introduction to the AdS/CFT correspondence}},
  \href{https://doi.org/10.1007/978-3-319-12238-0_10}{\emph{Springer Proc.
  Phys.} {\bfseries 161} (2015) 411}
  [\href{https://arxiv.org/abs/1310.4319}{{\ttfamily 1310.4319}}].

\bibitem{VanRaamsdonk:2016exw}
M.~Van~Raamsdonk, \emph{{Lectures on Gravity and Entanglement}},  in
  \emph{{Proceedings, Theoretical Advanced Study Institute in Elementary
  Particle Physics: New Frontiers in Fields and Strings (TASI 2015): Boulder,
  CO, USA, June 1-26, 2015}}, pp.~297--351, 2017,
  \href{https://arxiv.org/abs/1609.00026}{{\ttfamily 1609.00026}},
  \href{https://doi.org/10.1142/9789813149441_0005}{DOI}.

\bibitem{PhysRevD.14.2460}
S.~W. Hawking, \emph{Breakdown of predictability in gravitational collapse},
  \href{https://doi.org/10.1103/PhysRevD.14.2460}{\emph{Phys. Rev. D}
  {\bfseries 14} (1976) 2460}.

\bibitem{Mathur:2009hf}
S.~D. Mathur, \emph{{The Information paradox: A Pedagogical introduction}},
  \href{https://doi.org/10.1088/0264-9381/26/22/224001}{\emph{Class. Quant.
  Grav.} {\bfseries 26} (2009) 224001}
  [\href{https://arxiv.org/abs/0909.1038}{{\ttfamily 0909.1038}}].

\bibitem{Polchinski:2016hrw}
J.~Polchinski, \emph{{The Black Hole Information Problem}},  in
  \emph{{Proceedings, Theoretical Advanced Study Institute in Elementary
  Particle Physics: New Frontiers in Fields and Strings (TASI 2015): Boulder,
  CO, USA, June 1-26, 2015}}, pp.~353--397, 2017,
  \href{https://arxiv.org/abs/1609.04036}{{\ttfamily 1609.04036}},
  \href{https://doi.org/10.1142/9789813149441_0006}{DOI}.

\bibitem{Stoica:2018uli}
O.~C. Stoica, \emph{{Revisiting the black hole entropy and the information
  paradox}}, \href{https://doi.org/10.1155/2018/4130417}{\emph{Adv. High Energy
  Phys.} {\bfseries 2018} (2018) 4130417}
  [\href{https://arxiv.org/abs/1807.05864}{{\ttfamily 1807.05864}}].

\bibitem{Hayden:2007cs}
P.~Hayden and J.~Preskill, \emph{{Black holes as mirrors: Quantum information
  in random subsystems}},
  \href{https://doi.org/10.1088/1126-6708/2007/09/120}{\emph{JHEP} {\bfseries
  09} (2007) 120} [\href{https://arxiv.org/abs/0708.4025}{{\ttfamily
  0708.4025}}].

\bibitem{Sekino:2008he}
Y.~Sekino and L.~Susskind, \emph{{Fast Scramblers}},
  \href{https://doi.org/10.1088/1126-6708/2008/10/065}{\emph{JHEP} {\bfseries
  10} (2008) 065} [\href{https://arxiv.org/abs/0808.2096}{{\ttfamily
  0808.2096}}].

\bibitem{Lashkari:2011yi}
N.~Lashkari, D.~Stanford, M.~Hastings, T.~Osborne and P.~Hayden, \emph{{Towards
  the Fast Scrambling Conjecture}},
  \href{https://doi.org/10.1007/JHEP04(2013)022}{\emph{JHEP} {\bfseries 04}
  (2013) 022} [\href{https://arxiv.org/abs/1111.6580}{{\ttfamily 1111.6580}}].

\bibitem{Witten:1998zw}
E.~Witten, \emph{{Anti-de Sitter space, thermal phase transition, and
  confinement in gauge theories}},
  \href{https://doi.org/10.4310/ATMP.1998.v2.n3.a3}{\emph{Adv. Theor. Math.
  Phys.} {\bfseries 2} (1998) 505}
  [\href{https://arxiv.org/abs/hep-th/9803131}{{\ttfamily hep-th/9803131}}].

\bibitem{Maldacena:2001kr}
J.~M. Maldacena, \emph{{Eternal black holes in anti-de Sitter}},
  \href{https://doi.org/10.1088/1126-6708/2003/04/021}{\emph{JHEP} {\bfseries
  04} (2003) 021} [\href{https://arxiv.org/abs/hep-th/0106112}{{\ttfamily
  hep-th/0106112}}].

\bibitem{hawking1982}
S.~W. Hawking and D.~N. Page, \emph{Thermodynamics of black holes in anti-de
  sitter space}, {\emph{Comm. Math. Phys.} {\bfseries 87} (1982) 577}.

\bibitem{Hubeny:2009rc}
V.~E. Hubeny, D.~Marolf and M.~Rangamani, \emph{{Hawking radiation from AdS
  black holes}},
  \href{https://doi.org/10.1088/0264-9381/27/9/095018}{\emph{Class. Quant.
  Grav.} {\bfseries 27} (2010) 095018}
  [\href{https://arxiv.org/abs/0911.4144}{{\ttfamily 0911.4144}}].

\bibitem{Rocha:2008fe}
J.~V. Rocha, \emph{{Evaporation of large black holes in AdS: Coupling to the
  evaporon}}, \href{https://doi.org/10.1088/1126-6708/2008/08/075}{\emph{JHEP}
  {\bfseries 08} (2008) 075} [\href{https://arxiv.org/abs/0804.0055}{{\ttfamily
  0804.0055}}].

\bibitem{Rocha:2009xy}
J.~V. Rocha, \emph{{Evaporation of large black holes in AdS: Greybody factor
  and decay rate}},
  \href{https://doi.org/10.1088/1126-6708/2009/08/027}{\emph{JHEP} {\bfseries
  08} (2009) 027} [\href{https://arxiv.org/abs/0905.4373}{{\ttfamily
  0905.4373}}].

\bibitem{Penington:2019npb}
G.~Penington, \emph{{Entanglement Wedge Reconstruction and the Information
  Paradox}},  \href{https://arxiv.org/abs/1905.08255}{{\ttfamily 1905.08255}}.

\bibitem{Almheiri:2019psf}
A.~Almheiri, N.~Engelhardt, D.~Marolf and H.~Maxfield, \emph{{The entropy of
  bulk quantum fields and the entanglement wedge of an evaporating black
  hole}}, \href{https://doi.org/10.1007/JHEP12(2019)063}{\emph{JHEP} {\bfseries
  12} (2019) 063} [\href{https://arxiv.org/abs/1905.08762}{{\ttfamily
  1905.08762}}].

\bibitem{Almheiri:2020cfm}
A.~Almheiri, T.~Hartman, J.~Maldacena, E.~Shaghoulian and A.~Tajdini,
  \emph{{The entropy of Hawking radiation}},
  \href{https://arxiv.org/abs/2006.06872}{{\ttfamily 2006.06872}}.

\bibitem{Freedman:1999gp}
D.~Freedman, S.~Gubser, K.~Pilch and N.~Warner, \emph{{Renormalization group
  flows from holography supersymmetry and a c theorem}},
  \href{https://doi.org/10.4310/ATMP.1999.v3.n2.a7}{\emph{Adv. Theor. Math.
  Phys.} {\bfseries 3} (1999) 363}
  [\href{https://arxiv.org/abs/hep-th/9904017}{{\ttfamily hep-th/9904017}}].

\bibitem{deBoer:2000cz}
J.~de~Boer, \emph{{The Holographic renormalization group}},
  \href{https://doi.org/10.1002/1521-3978(200105)49:4/6<339::AID-PROP339>3.0.CO;2-A}{\emph{Fortsch.
  Phys.} {\bfseries 49} (2001) 339}
  [\href{https://arxiv.org/abs/hep-th/0101026}{{\ttfamily hep-th/0101026}}].

\bibitem{Rangamani:2016dms}
M.~Rangamani and T.~Takayanagi, \emph{{Holographic Entanglement Entropy}},
  \href{https://doi.org/10.1007/978-3-319-52573-0}{\emph{Lect. Notes Phys.}
  {\bfseries 931} (2017) pp.1}
  [\href{https://arxiv.org/abs/1609.01287}{{\ttfamily 1609.01287}}].

\bibitem{Leichenauer:2014nxa}
S.~Leichenauer, \emph{{Disrupting Entanglement of Black Holes}},
  \href{https://doi.org/10.1103/PhysRevD.90.046009}{\emph{Phys. Rev.}
  {\bfseries D90} (2014) 046009}
  [\href{https://arxiv.org/abs/1405.7365}{{\ttfamily 1405.7365}}].

\bibitem{Brown:2018kvn}
A.~R. Brown, H.~Gharibyan, A.~Streicher, L.~Susskind, L.~Thorlacius and
  Y.~Zhao, \emph{{Falling Toward Charged Black Holes}},
  \href{https://doi.org/10.1103/PhysRevD.98.126016}{\emph{Phys. Rev.}
  {\bfseries D98} (2018) 126016}
  [\href{https://arxiv.org/abs/1804.04156}{{\ttfamily 1804.04156}}].

\bibitem{Myers:2010xs}
R.~C. Myers and A.~Sinha, \emph{{Seeing a c-theorem with holography}},
  \href{https://doi.org/10.1103/PhysRevD.82.046006}{\emph{Phys. Rev. D}
  {\bfseries 82} (2010) 046006}
  [\href{https://arxiv.org/abs/1006.1263}{{\ttfamily 1006.1263}}].

\bibitem{Myers:2010tj}
R.~C. Myers and A.~Sinha, \emph{{Holographic c-theorems in arbitrary
  dimensions}}, \href{https://doi.org/10.1007/JHEP01(2011)125}{\emph{JHEP}
  {\bfseries 01} (2011) 125} [\href{https://arxiv.org/abs/1011.5819}{{\ttfamily
  1011.5819}}].

\bibitem{1983GReGr..15..195U}
W.~G. {Unruh} and R.~M. {Wald}, \emph{{How to mine energy from a black hole.}},
  \href{https://doi.org/10.1007/BF00759206}{\emph{General Relativity and
  Gravitation} {\bfseries 15} (1983) 195}.

\bibitem{Lawrence:1993sg}
A.~E. Lawrence and E.~J. Martinec, \emph{{Black hole evaporation along
  macroscopic strings}},
  \href{https://doi.org/10.1103/PhysRevD.50.2680}{\emph{Phys. Rev. D}
  {\bfseries 50} (1994) 2680}
  [\href{https://arxiv.org/abs/hep-th/9312127}{{\ttfamily hep-th/9312127}}].

\bibitem{Frolov:2000kx}
V.~P. Frolov and D.~Fursaev, \emph{{Mining energy from a black hole by
  strings}}, \href{https://doi.org/10.1103/PhysRevD.63.124010}{\emph{Phys. Rev.
  D} {\bfseries 63} (2001) 124010}
  [\href{https://arxiv.org/abs/hep-th/0012260}{{\ttfamily hep-th/0012260}}].

\bibitem{Brown:2012un}
A.~R. Brown, \emph{{Tensile Strength and the Mining of Black Holes}},
  \href{https://doi.org/10.1103/PhysRevLett.111.211301}{\emph{Phys. Rev. Lett.}
  {\bfseries 111} (2013) 211301}
  [\href{https://arxiv.org/abs/1207.3342}{{\ttfamily 1207.3342}}].

\bibitem{Mistry:2017ubm}
R.~Mistry, S.~Upadhyay, A.~F. Ali and M.~Faizal, \emph{{Hawking radiation power
  equations for black holes}},
  \href{https://doi.org/10.1016/j.nuclphysb.2017.08.010}{\emph{Nucl. Phys. B}
  {\bfseries 923} (2017) 378}
  [\href{https://arxiv.org/abs/1709.01163}{{\ttfamily 1709.01163}}].

\bibitem{Page:2015rxa}
D.~N. Page, \emph{{Finite upper bound for the Hawking decay time of an
  arbitrarily large black hole in anti--de Sitter spacetime}},
  \href{https://doi.org/10.1103/PhysRevD.97.024004}{\emph{Phys. Rev. D}
  {\bfseries 97} (2018) 024004}
  [\href{https://arxiv.org/abs/1507.02682}{{\ttfamily 1507.02682}}].

\bibitem{Ford:1995gb}
L.~Ford and T.~A. Roman, \emph{{Averaged energy conditions and evaporating
  black holes}}, \href{https://doi.org/10.1103/PhysRevD.53.1988}{\emph{Phys.
  Rev. D} {\bfseries 53} (1996) 1988}
  [\href{https://arxiv.org/abs/gr-qc/9506052}{{\ttfamily gr-qc/9506052}}].

\bibitem{Lesourd}
M.~Lesourd, \emph{A remark on the energy conditions for hawking's area
  theorem}, \href{https://doi.org/10.1007/s10714-018-2377-1}{\emph{General
  Relativity and Gravitation} {\bfseries 50} (2018) 61}.

\bibitem{McGough:2016lol}
L.~McGough, M.~Mezei and H.~Verlinde, \emph{{Moving the CFT into the bulk with
  $ T\overline{T} $}},
  \href{https://doi.org/10.1007/JHEP04(2018)010}{\emph{JHEP} {\bfseries 04}
  (2018) 010} [\href{https://arxiv.org/abs/1611.03470}{{\ttfamily
  1611.03470}}].

\bibitem{Israel1966}
W.~Israel, \emph{Singular hypersurfaces and thin shells in general relativity},
  \href{https://doi.org/10.1007/BF02710419}{\emph{Il Nuovo Cimento B
  (1965-1970)} {\bfseries 44} (1966) 1}.

\bibitem{poisson_2004}
E.~Poisson, \emph{Hypersurfaces}, p.~59–117.
\newblock Cambridge University Press, 2004.
\newblock 10.1017/CBO9780511606601.005.

\end{thebibliography}\endgroup
\bibliographystyle{JHEP}



\end{document}